\documentclass{aa}  

\usepackage{graphicx}
\usepackage{txfonts}

\usepackage[normalem]{ulem}
\usepackage{natbib}
\bibpunct{(}{)}{;}{a}{}{,}
\usepackage{lscape}
\usepackage{xcolor}
\usepackage{url}

\usepackage[colorlinks=true,
    linkcolor=blue,
    citecolor=blue,
    filecolor=magenta,      
    urlcolor=cyan]{hyperref}
\usepackage{subfigure}

\begin{document}

   \title{Testing compact massive black hole binary candidates through multi-epoch spectroscopy}

   \author{Lorenzo Bertassi
          \inst{1,2,3} \fnmsep\thanks{l.bertassi@campus.unimib.it}
          \and
          Erika Sottocorno\inst{1}
          \and 
          Fabio Rigamonti\inst{2,3,4}
          \and
          Daniel D'Orazio\inst{5,6,7}
          \and
          Michael Eracleous \inst{8}
          \and \\
          Zolt\'an Haiman\inst{9,10,11}
          \and
          Massimo Dotti\inst{1,2,3}
          }

   \institute{Università degli Studi di Milano-Bicocca, Piazza della Scienza 3, 20126 Milano, Italy
    \and
    INAF - Osservatorio Astronomico di Brera, via Brera 20, I-20121 Milano, Italy
    \and
    INFN, Sezione di Milano-Bicocca, Piazza della Scienza 3, I-20126 Milano, Italy
    \and
    Como Lake centre for AstroPhysics (CLAP), DiSAT, Università dell’Insubria, via Valleggio 11, 22100 Como, Italy
    \and
    Space Telescope Science Institute, 3700 San Martin Drive, Baltimore, MD 21218, USA
    \and
    Department of Physics and Astronomy, Johns
    Hopkins University, 3400 North Charles Street, Baltimore,
    Maryland 21218, USA
    \and
    Niels Bohr International Academy, Niels Bohr Institute, Blegdamsvej 17, DK-2100 Copenhagen Ø, Denmark
    \and
    Department of Astronomy and Astrophysics, Penn State University, 525 Davey Lab, 251 Pollock Road, University Park, PA 16802, USA
    \and
    Department of Astronomy, Columbia University, New York, NY 10027, USA
    \and 
    Department of Physics, Columbia University, New York, NY 10027 USA
    \and 
    Institute of Science and Technology Austria (ISTA), Am Campus 1, Klosterneuburg 3400, Austria
    }
   \titlerunning{Multi-epoch spectral test for MBHB candidates}
   \authorrunning{Bertassi et al.}

   \date{Received ...; accepted...}

  \abstract{
  Emission from two massive black holes (MBHs) bound in a close binary is expected to be modulated by different processes, such as the Doppler boost due to the orbital motion, accretion rate variability generated by the interaction with a circumbinary disc, and binary gravitational self-lensing. When the binary is compact enough, the two black holes are thought to be surrounded by a common broad-line region that reprocesses the impinging periodically varying ionising flux, creating broad emission lines with variable line shapes. Therefore, the study of broad emission line variability through multi-epoch spectroscopic campaigns is of paramount importance for the unambiguous identification of a binary. In this work, we study the response of a disc-like broad-line region to the Doppler-boosted ionising flux emitted by sub-milliparsec MBH binaries on a circular orbit and compare it with the response of a broad-line region illuminated by a single MBH with a periodically but isotropically varying intrinsic luminosity.
  We show that in the binary case, the time lags of the blue and red wings of the broad emission lines, arising from diametrically opposite sides of the circumbinary disc, are out of phase by half of the binary's orbital period, as they each respond to the periodic ‘lighthouse’ modulation from the binary's continuum emission.  This asymmetric time lag represents a new binary signature that cannot be mimicked by a single MBH.
}

   \maketitle

\section{Introduction}\label{sec:introduction}

Massive black hole binaries (MBHBs) are predicted to form in the aftermath of galaxy mergers \citep{Begelman}. These objects are expected to be among the loudest gravitational wave (GW) sources that will be identified by future space-borne interferometers \citep[e.g.][]{LISA} and pulsar timing array campaigns \citep{Verbiest2016}. Recently, evidence of a GW background, possibly produced by supermassive black hole binaries, has been detected by PTA experiments \citep[see, e.g.][]{Nanograv15, EPTA2023}. Identifying these sources simultaneously through electromagnetic (EM) observations will increase our knowledge of black hole inspirals, accretion physics, cosmology, and general relativity \citep[see, e.g. ][]{Baker-decadal, De_Rosa_dual_AGN, LISA}. 

While there are confirmed EM detections of dual active galactic nuclei (AGNs) at $\sim$ kiloparsec-parsec separations (see \citealt{De_Rosa_dual_AGN} for a review, \citealt{Trindade24}  for a recent candidate with a separation of 100 pc and \citealt{Rodriguez_2006} for a candidate with a separation of 7.3 pc), strong evidence of gravitationally bound MBHBs at sub-parsec scales is still missing. For this reason, several indirect observational signatures have been proposed \citep[see e.g.][for recent discussions]{Nguyen-Bogdanovic_1, Dust_lighthouse, Nguyen-Bogdanovic_2, Fast_test, EM_signatures}.

Among these signatures, the photometric ones are of particular interest as they will be searched for in forthcoming large observational campaigns, such as the Vera Rubin Observatory's 10-year Legacy Survey of Space and Time \citep[LSST, see][]{LSST} and the Roman Space Telescope’s High Latitude Time Domain Survey \citep[HLTDS, see][]{Roman},  with the appropriate depth and cadence. Examples of these signatures are the variability in the continuum emission caused by the interaction between the binary and a circumbinary disc \citep[see][]{Accretion_variability} and by relativistic effects, such as Doppler boosting \citep[see][]{Doppler_boost} and gravitational self-lensing \citep[see][]{Self_lensing}.

The length and cadence of these new time domain surveys will allow us to reject sources that present false periodicities due to red noise observed in optical and UV light curves \citep[e.g.][]{False_periodicies} for short period binaries, with $\rm{P} \lesssim 1 \ \rm{month}$ corresponding to separations of $a\sim 10^{-4} \ (M/10^6 \rm{M_{\odot})}^{1/3} \ \mathrm{pc} $, where $M$ is the binary total mass. Even though a fraction of false candidates can be rejected through long-term monitoring, the effective presence of a binary needs to be tested in detail as single massive black holes (MBHs) may have a variable intrinsic luminosity mimicking modulation in the light curves expected from binaries \citep[see e.g. the discussion in][]{Sandrinelli16}. Our work proposes a new spectroscopic test to prove the binary nature of photometrically identified sub-milliparsec MBHB candidates.
At such small separations, the hypothetical binaries will have mildly relativistic orbital speeds, approaching $0.1c$ for the more massive and compact systems, and are thought to be embedded in a common circumbinary broad-line region (BLR). If at least one of the two MBHs is accreting, the periodically modulated ionising flux impinging on the BLR is expected to affect the broad emission line (BEL) shape in unique ways, due to the beamed anisotropic irradiation. The BEL time evolution then encodes information about the presence of a binary.

In our work, we assume disc-like BLRs, and we model both single and double-peaked BELs by changing the inclination of the system. Non-asymmetric BELs can be modelled by including deviations from axisymmetry in the BLR emissivity profile \citep[see][]{Disc_BLR}. Under this assumption, we compute how the BEL responds to the Doppler-modulated ionising continuum, which varies to first order as $O(v/c)$, in the case of an MBHB, as well as to an intrinsically isotropically variable continuum emitted by a single MBH that has the exact same photometric modulation as is observed in the MBHB case.
The following discussion is tailored for the broad $\rm{H{\beta}}$ line, but the proposed test can be readapted to other optical-UV BELs by adjusting the ‘characteristic radius’ of the BLR. Our scenario has some similarities with the model proposed by \cite{Hutchings1977} to explain the 3-hour periodic variation
of the observed magnitudes and emission lines 
in the Nova VI500 Cyg. Unlike our model, in which the modulation is caused by the anisotropic emission due to the Doppler boost, the periodic emission is explained by the varying presentation aspect of a hot spot in a conventional cataclysmic variable model \citep[see][]{Robinson1976}. In their model, the nebula, which plays a role similar to that of the BLR in our work, responds to the varying impinging continuum and presents a multi-component, axisymmetric structure consisting of a disc and two polar blobs. Similarly to the binary signature that we are proposing (see Section~\ref{sec:results}), the observed spectroscopic features of the nova emission lines cannot be explained under the assumption of an isotropic source.

In Section~\ref{sec:methodology}, we describe the underlying assumptions of our model, the methods with which the BELs are constructed, and how the fluxes of interest are computed. In Section~\ref{sec:results}, we illustrate our main results. Finally, in Section~\ref{sec:conclusions}, we draw our main conclusions and propose the next steps to generalise our work to more complex scenarios.

\section{Methodology}\label{sec:methodology}
In our work, we compare the response of a disc-like BLR (detailed in Section~\ref{subsec:BLR}) to the flux emitted by either ($i$) two MBHs bound in a close binary, whose properties are detailed in Section~\ref{subsec:binary}, under the assumption that each MBH has a constant luminosity and a radiation pattern shaped by Doppler boosting, or ($ii$) a single MBH at rest at the centre of the BLR, with a time-dependent luminosity such that the intrinsic emission in the continuum would be indistinguishable from the binary case (see Section~\ref{subsec:BEL_construction}). 
The difference between the binary and single MBH arises because the binary's emission is beamed and anisotropic, illuminating different parts of the BLR at different times, analogous to a ‘lighthouse’~\citep[][]{Dust_lighthouse}.
To assess whether this allows us to distinguish binaries from single MBHs, we computed the BEL shape generated by the impinging flux in the two scenarios (see Section \ref{subsec:BEL_construction}), we found the light curves from the red and blue components of the BEL, as is discussed in Section \ref{subsec:light_curves}, and finally we searched for the delays of the red and blue light curves relative to the ionising continuum. The last step was done by cross-correlating the three light curves using PyCCF \citep{pythoncode18}\footnote{\url{http://ascl.net/code/v/1868}}, the Python version of the original algorithm discussed in  \cite{PYCCF}. This algorithm allows one to search for time lags between unevenly sampled light curves through linear interpolation. Also, the code allows one to compute the uncertainties associated with the time lags through Monte Carlo iterations. The test we propose is similar to the one discussed in \cite{Dust_lighthouse}, in which the (photometric) response of the torus is modelled for MBHBs as well as for single MBHs with modulated continua. The most important difference between the two studies is that our analysis can distinguish the response of different parts of the reprocessing structure (the BLR, in our scenario), encoded in the spectral profile of the BELs, as is detailed below.

In what follows, we consider stochastic noise caused by intrinsic AGN variability
as discussed in \ref{sec:results}, and we implicitly assume the limit of infinite signal-to-noise ratio (S/N) in the observed spectra. This last assumption will be relaxed in future analyses.

\subsection{Binaries and Doppler boosting}\label{subsec:binary}
In the binary scenario, we model a binary orbiting within a common BLR.
Even if from hydrodynamical simulations, binaries are expected to have a certain degree of eccentricity \citep[e.g.][]{Roedig11, Munoz19, Zrake21, D'Orazio_Duffell21} depending on their mass ratio \citep[see][]{Siwek}, in our work, we start by assuming circular binaries. The effect of binary eccentricity will be explored in a follow-up study. 

\begin{figure}
    \centering
    \includegraphics[width=\hsize]{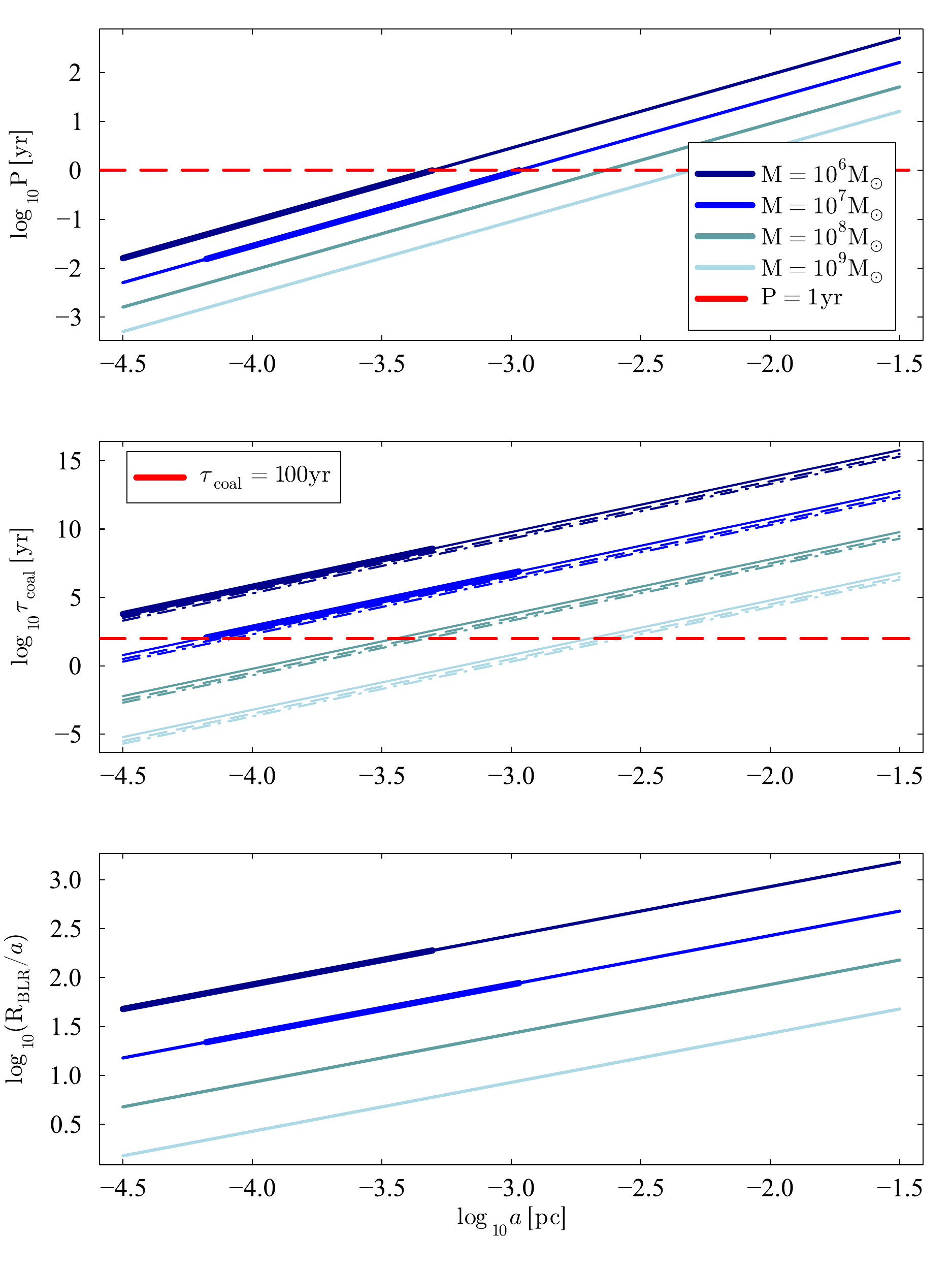}
    \caption{Upper panel: Binary period for binaries with different total mass and separations. The horizontal dashed red line indicates where the period is $\rm{P}=1 \ \rm{yr}$. Middle panel: Coalescence time of the binaries with different masses (see the colours in the upper panel) and mass ratios, $q$. The solid lines show binaries with $q=0.1$, dashed lines show binaries with $q=0.25$, and finally, dash-dotted lines show binaries with $q=1$. The horizontal dashed red line indicates where the coalescence time due to GW emission is $\tau_{\rm{coal}}=100 \ \rm{yr}$. Lower panel: Ratio between the characteristic radius of the BLR and the binary separation for different masses. In all three panels, the segments of the lines that meet the conditions of interest in this work (binaries with total masses of around $10^6-10^7 \rm{M_{\odot}}$, an orbital period of $\rm{P}<1 \ \rm{yr}$, and a coalescence time due to GW emission of $\tau_{\rm{coal}}> 100 \ \rm{yr}$) are drawn with a thicker line. In the middle panel, only the case of $q=0.1$ has a thick segment to maintain readability. }
    \label{fig:multiplot}
\end{figure}

We started by exploring the binary parameter space to identify the binaries that were most likely to be diagnosed with our proposed method. In the upper panel of Figure \ref{fig:multiplot}, we show the orbital period, $P$, of the binary. Looking at the source for multiple periods is important as it can prune out contaminants from intrinsic red noise AGN variability. As can be seen, at separations smaller than $\sim 10^{-3} \ \mathrm{pc}$ the period is shorter than a year for most binaries.

In the middle panel of Figure \ref{fig:multiplot}, we show the coalescence time through GW emission, $\tau_{\rm{coal}}$, of the binary at a given initial separation for different masses and mass ratios. This quantity is defined as \citep[][]{maggiore_GW}
\begin{equation}
    \tau_{\rm{coal}}=\frac{5}{256} \frac{c^5 a^4}{G^3 M^2 \mu},
\end{equation}
with $\mu$ and $a$ being, respectively, the reduced mass and the orbital separation of the binary. From this plot, it becomes clear that at separations of a megaparsec or smaller, binaries with a high total mass will be rare in observations as they will coalesce quickly. Also, the majority of AGNs that LSST will observe will be at their faint detection threshold, powered by lower-mass ($M\sim 10^6~{\rm M_\odot}$) MBHs \citep[see][]{Xin_forecast}. For these reasons, we focus on lighter $(10^6-10^7 \rm{M_{\odot}})$ binaries.

The emission from the two MBHs is assumed to be produced by a Novikov-Thorne disc \citep[see][]{Novikov_Thorne, Page1974}\footnote{The viscosity parameter of both discs is set to $\alpha=0.01$}, with the discs around the MBHs having an outer radius given by the Roche lobe radius. Considering the secondary MBH, this radius is well approximated by \citep[see][]{Roche_lobes}
\begin{equation}\label{eq:Roche_lobes}
    R_{\mathrm{RL},2}\approx 0.49 a \frac{q^{\frac{2}{3}}}{0.6 q^{\frac{2}{3}} + \ln(1+q^{\frac{1}{3}})}
,\end{equation}
where $q\equiv M_2/M_1\leq 1$ is the binary mass ratio. However, for very thin, close-to-Keplerian discs, the outer radius is smaller by a factor of 4-5  \citep[see][]{Truncation_radius}. 

The luminosity emitted from the two discs depends on the mass accretion rate of the two MBHs, which is related to the binary mass ratio \citep[see][]{m_dot_ratio}. Expressing the accretion rate of the two MBHs in terms of their individual Eddington ratios, $f_{\mathrm{Edd}}\propto \eta\dot{M}/M$, one finds
\begin{equation}
    \frac{f_1}{f_2}=\frac{\eta_1}{\eta_2}q\left(0.1+0.9 \ q\right)
    \label{eq:Edd_ratio_frac}    
,\end{equation}
where $f_1$ and $f_2$ are the Eddington ratios and $\eta_1$ and $\eta_2$ are the radiative efficiencies of the primary and secondary MBH. Due to the limited constraints on the distribution of MBH spins in the following, we assume, arbitrarily, the normalised angular momentum to be $\Tilde{a}=0.8 J/Mc$ for both MBHs and that the accretion disc is co-rotating with the MBH. Thus, the radiative efficiency is $\eta \approx 0.12$. 

Since the two MBHs are moving, their emission is subjected to Doppler boosting. The MBH rest-frame frequency, $\nu$, of a photon emitted by its accretion disc is related to the frequency, $\nu^{\prime}$,  seen by an observer as $\nu^{\prime} /\nu=D$, where $D$ is the Doppler coefficient. This quantity can be written in terms of the relative velocity between the emitting source and the observer as 
\begin{equation}\label{eq:Doppler_coeff}
    D=\frac{1}{\gamma\ (1-\beta \ \cos \theta)},
\end{equation}
where $\theta$ is the angle defined in Equation~(\ref{eq: angle_v_BLR}) below,  $\beta\equiv(\vec{\upsilon}_{\rm{em}} - \vec{\upsilon}_{\rm el})/c$ is the relative velocity between the emitter and the BLR element normalised to the speed of light, and finally $\gamma\equiv1/\sqrt{1-\beta^2}$ is the Lorentz factor. In $\beta$, the emitter's velocity direction is not the one seen at time $t$ but the direction of motion at the retarded time of emission defined in Equation~(\ref{eq:time_prime}) below. 
The Doppler boosted ionising flux is given by
\begin{equation}\label{eq:boosted_integral}
    F^{\rm{boost}}_{\rm{ion}}=D^4 \int_{\nu_{\rm{ion}}/D}^\infty F_{\nu/D} \ d\nu 
,\end{equation}
where $h \nu_{\rm{ion}}=1 \ \rm{Ry}$ and $\nu$ is the rest-frame frequency.

\begin{figure}[t]

    \subfigure{
        \includegraphics[width=4cm]{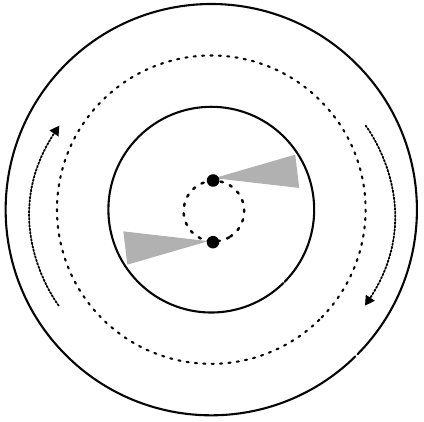}
        }
    \subfigure{
        \includegraphics[width=4 cm]{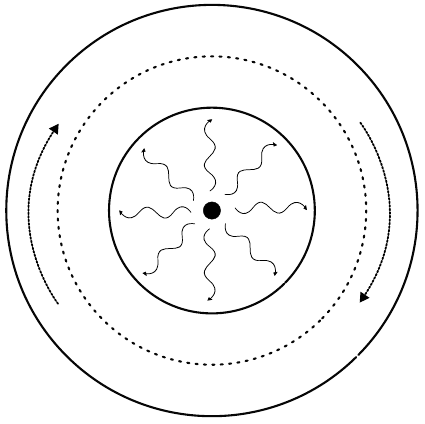}
        }
    \caption{Illustration of the emission patterns in two scenarios. In the binary scenario (left), the emission from the accretion disc is beamed along the direction of motion, while in the single MBH scenario (right) the emission is the same in all directions.}
    \label{fig:qualitative_schemes}
\end{figure}

In Figure \ref{fig:qualitative_schemes}, we illustrate the emission pattern in the binary and single MBH scenarios. In the single MBH case, the emission is isotropic, while in the MBHB scenario, the orbital motion beams the emission along the direction of motion. Since the orbital motion of the binary is periodic, the variation in the ionising flux seen by a given observer (or fluid element of the BLR in our case) is expected to be periodic, as well \citep[\textit{e.g.},][]{Doppler_boost}.

\subsection{Broad-line region}\label{subsec:BLR}

\begin{figure}[h!]
    \centering
    \includegraphics[width=0.9\hsize]{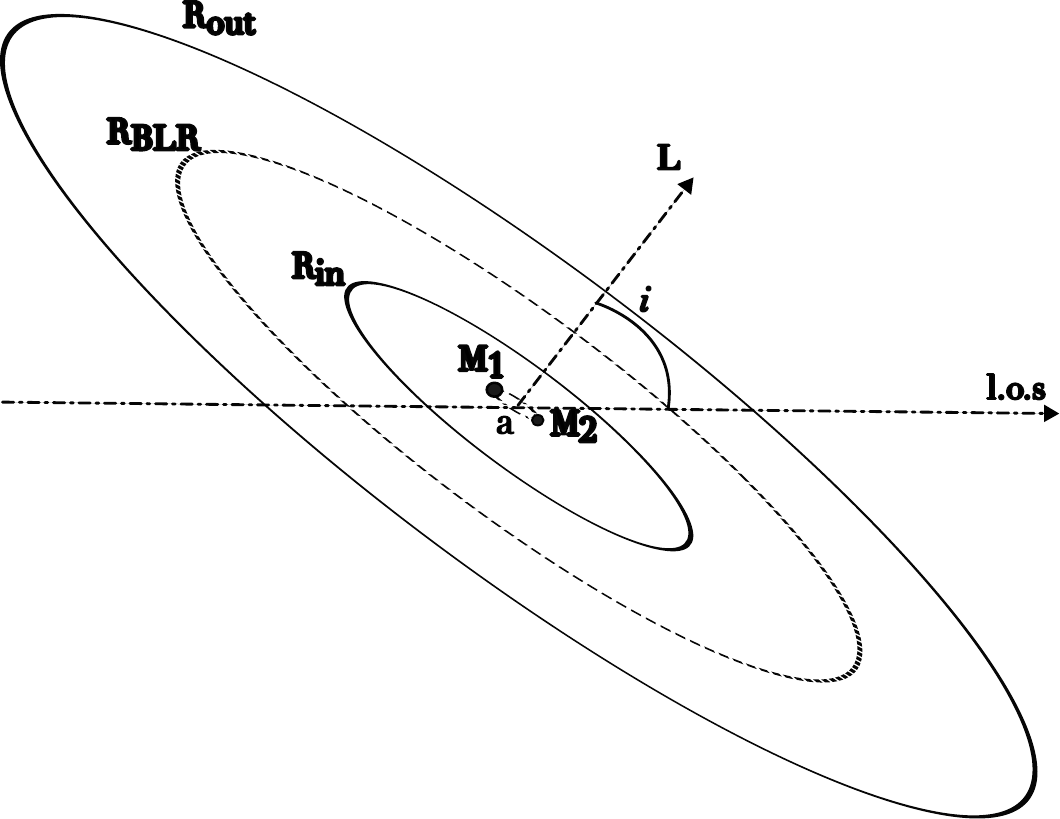}
    \caption{Illustration of the BLR geometry. Here $R_{\mathrm{in}}$ and $R_{\mathrm{out}}$ are the inner and outer radii of the BLR, while $R_{\mathrm{BLR}}$ is its characteristic radius. $\textbf{L}$ is the angular momentum of the system, l.o.s is the line of sight to an observer, and finally $i$ is the inclination angle between $\textbf{L}$ and the line of sight. $M_1$ and $M_2$ are the masses of the two MBHs, while $a$ is the binary separation. The binary separation and the BLR characteristic radius are not drawn to scale.}
    \label{fig:BLR_geom}
\end{figure}

In this work, the BLR that surrounds the binary has a disc-like geometry\footnote{Such an assumption can explain within a single model the emission of single-peaked and double-peaked BELs \citep[e.g.][]{DPE_Eracleous09}, as well as the observed inclination dependence of the BEL line in radio-loud AGN \citep[e.g.][]{Wills86, McLure01, Runnoe13}}. The BLR disc and the binary are assumed to be coplanar for two reasons: $(i)$ because in gas-rich galaxy mergers both the gas flowing towards the centre of the galaxy remnant and the two MBHs binding in a binary are expected to be affected by the angular momentum of the merger, so that the binary is expected to form already within a coplanar gas reservoir \citep[e.g.][]{CapeloDotti17}, and $(ii)$ because even if a misaligned binary-BLR forms, gravitational torques are expected to realign the two structures on $\sim 1-10$ Myr timescales \citep[e.g.][]{MillerKrolik13}. We nevertheless briefly comment on the impact of a relative MBHB-BLR misalignment at the end of section~\ref{subsec:light_curves}. The whole MBHB-BLR system is inclined with an angle, $i$, relative to the line of sight as shown in Figure~\ref{fig:BLR_geom}.

The characteristic radius of the BLR is defined \citep[see][]{Kaspi00, Bentz, Fast_test} as 
\begin{equation}
    R_\mathrm{BLR}= c \ \tau_{\rm LTT} =11 \left(f_{\rm{Edd}}\frac{M}{10^6 M_{\odot}}\right)^{0.519} \ \rm{ld},
    \label{eq:emiss_weighted_radius}
\end{equation} 
where $\tau_{\rm{LTT}}$ is the light travel time across the BLR. Here we assume that $\tau_{\rm{LTT}}$ is of the same order as the binary orbital period ($P$), $\sim 10 $ light-days in the example above. In such a case, we expect the response of the BEL profile to the orbit of the binary to be maximised. More specifically, if $P \ll \tau_{\rm LTT}$ the effect of the periodicity in the flux from the BLR would be averaged out over multiple periods, while if $P \gg \tau_{\rm LTT}$, the length of an observational campaign aiming to detect a modulation in the BEL profiles will increase, and, for too large separations of the MBHB, the assumption of having a single BLR around the two MBHs will no longer be valid \citep[see][]{Fast_test}.
In this work, the orbital period and the average light travel time are specifically assumed to be $\tau_{\rm LTT} = 0.3 \ P $.

The inner and outer radii are given by $R_{\mathrm{BLR,in}}=0.5 \ R_{\mathrm{BLR}}$ and $R_{\mathrm{BLR,out}}=1.5 \ R_{\mathrm{BLR}}$. Once the binary masses and separation are chosen, $R_{\rm{BLR}}$ is fixed by the value of $\tau_{\rm{LTT}}$ and the Eddington ratios for the two MBHs are set by combining Equations \ref{eq:Edd_ratio_frac} and \ref{eq:emiss_weighted_radius}. 
The resulting values for the Eddington ratios of the individual BHs fall between $0.01$ and $\approx 0.6$, which is consistent with the observations \citep[see][]{Shankar2013}.

In the lower panel of Figure \ref{fig:multiplot}, we show the ratio between the characteristic radius of the BLR and the binary separation. 
For lighter binaries ($10^6-10^7 \rm{M_{\odot}}$), which are also the binaries of interest for our work as discussed in section \ref{subsec:binary}, the orbital separation is always much smaller than the characteristic radius of the BLR. For this reason, as well as for simplicity, we assume the binaries to be point-like sources at the centre of the BLR. Under these assumptions, the light travel time of a photon emitted by either of the two MBHs to reach an element of the BLR at a distance, $r_{\mathrm{el}}$, will be well approximated by $r_{\mathrm{el}}/c$ and the effective emission time of a photon from the central source can be computed as in Equation (\ref{eq:time_prime}) below.

\begin{figure}[h!]
    \centering
    \includegraphics[width=0.8\hsize]{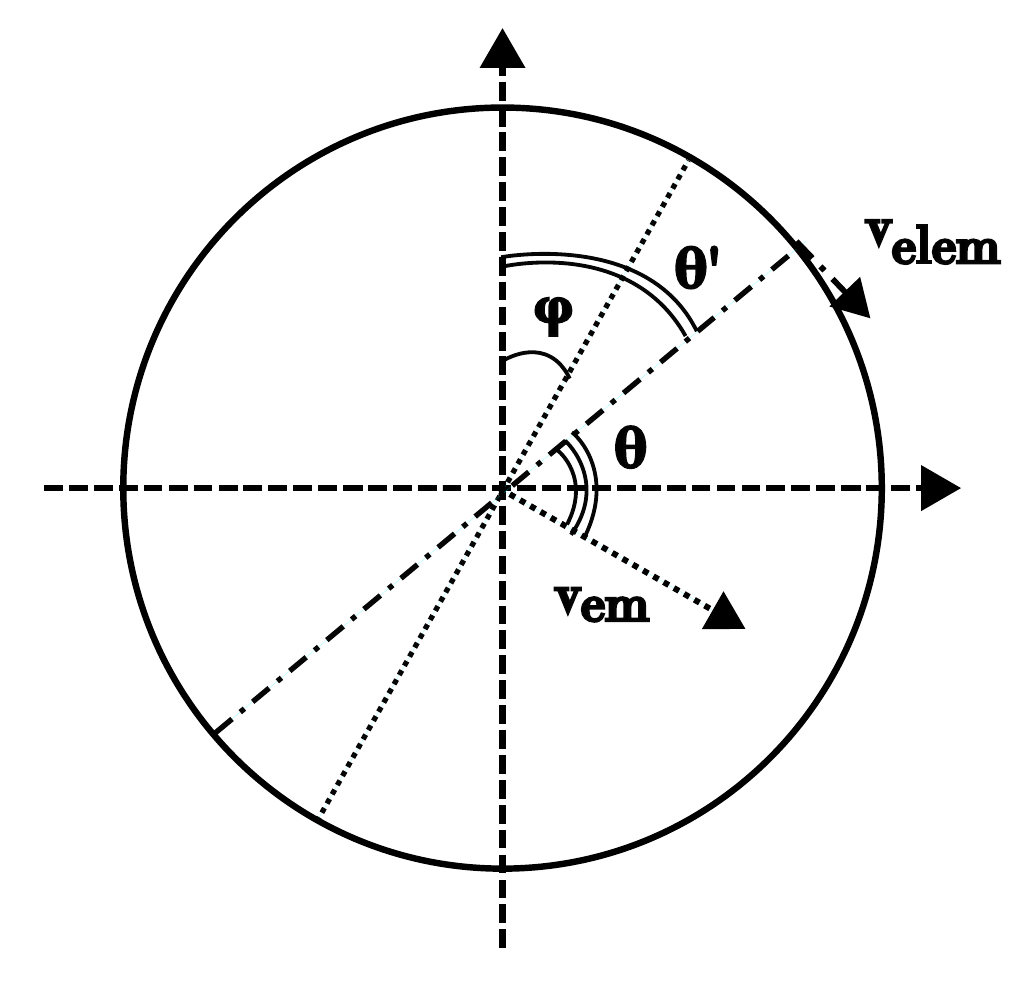}
    \caption{Geometry of the system seen as face-on. The rotation of the angles relative to the Cartesian co-ordinate system is clockwise.}
    \label{fig:face-on_geometry}
\end{figure}

Consider the Cartesian reference frame centred at the centre of the binary and also of the BLR shown in Figure \ref{fig:face-on_geometry}. Assume that the inclination of the binary relative to an observer at an arbitrarily large distance, $d_0$, is the same as the inclination of the BLR disc, $i$, i.e. that the binary's orbit is coplanar with its BLR disc, as is shown in Figure \ref{fig:BLR_geom}. The angle between the direction of motion, $\bold{v}_{\rm em}$, of an MBH and the direction towards a BLR element is given by
\begin{equation}\label{eq: angle_v_BLR}
    \theta_{1,2}=\pi/2-\theta^{\prime}+\varphi_{1,2} \ ,
\end{equation}
where $\theta^\prime$ is the azimuthal angle of the BLR element with respect to a reference direction (the positive $y$ axis in Fig.~\ref{fig:face-on_geometry})  and $\varphi_{1,2}$ is the azimuthal position of the MBH (offset by an angle $-\pi/2$ from its orbital velocity vector).

The light emitted by the central pointlike source will reach the observer after a time of $\sim d_0/c$, where $d_0$ is the distance between the observer and the centre of the system. Since the BLR disc can be inclined relative to the line of sight, the light emitted at any time by the MBH's accretion disc will reach the observer after being reprocessed by a BLR element with 2D co-ordinates (in the BLR plane), $r^\prime$ and $\theta^{\prime}$, after a time delay, $\Delta t$, defined as 
\begin{equation}
    \Delta t (r^\prime,\theta^{\prime})=t-\frac{r^\prime}{c} \ \left(1+\sin i\cos\theta^{\prime}\right) \ ,
    \label{eq:time_prime}
\end{equation}
where $t\equiv d_0/c$. The term $r^\prime \sin i\cos\theta^{\prime}/c$ comes from the effective distance of the element from the observer given by $d(r^\prime,\theta^{\prime})=d_{0}+r^\prime \sin i \cos\theta^{\prime}$.

Our BLR model is simplified by the following assumptions: 
($i$) In Equation \ref{eq:time_prime}, we are not taking into consideration the recombination time of the BLR, i.e. the time necessary for the line emission to respond to the new ionising continuum. The recombination time can be written as $t_{\mathrm{rec}}=(n\alpha_r)^{-1}$, where $n$ is the number density and $\alpha_r$ is the recombination coefficient. Given the lower limit to the density $n=10^{12} \ \rm{cm}^{-3}$ given by the detection of strong and broad collisionally excited resonance lines in the UV \citep[see][]{Moloney2014} and the typical BLR temperature ($T \sim  10^4 \ K$, resulting in a recombination coefficient, $\alpha_r \sim 10^{-13} \ \rm{cm}^3 \rm{s}^{-1}$, e.g. \citealt{Storey1995}), the recombination time is expected to be $t_{\rm{rec}}\lesssim 10 \ \rm{s}$, justifying our assumption;

($ii$) We do not consider any effect due to structural changes in the BLR.
Such changes would happen on the BLR dynamical time, $t_{\mathrm{dyn}}$, of the order of the period of a circular orbit at the BLR radius. Within the assumptions of our study, the ratio between $t_{\mathrm{dyn}}$ and the binary period is
$t_{\mathrm{dyn}}/P = (R_{\mathrm{BLR}}/a_\mathrm{bin})^{3/2}$,
ranging from a few thousand to tens of thousands for the binary parameters considered here. Although a perturbation exerted by the MBHB on scales much smaller than $R_{\rm{BLR}}$ could in principle propagate to the BLR, the details of such propagation would be model-dependent and are outside the limited scope of this study.

Instead of solving for the time-dependent ionisation equilibrium, we assume a phenomenological description for the emissivity profile, $\epsilon(\xi,\theta)$, which is able to reproduce the delays assuming the radius-luminosity relation discussed in \cite{Bentz}. In our model, the BLR emissivity profile features a spiral pattern superimposed on an otherwise axisymmetric disc-like BLR. This prescription, originally proposed by \cite{Storchi-Bergmann} and recently modified by 
\cite{sottocorno2025}, is a proxy of a broad family of perturbations that make some portion of the disc brighter than the rest of the disc. The axisymmetric term is instead modelled as a Gaussian centred at a radius proportional to $R_{\mathrm{BLR}}$ following the radius-luminosity relation. The assumed emissivity profile is discussed in more detail in Appendix \ref{app:BLR_emis}.

\subsection{BEL construction} \label{subsec:BEL_construction}

The BEL profile observed at any given time was computed as the sum of the contributions from all the elements of the BLR.  
Dividing the BLR into $\rm N_r$ radial and $\rm {N_{\theta^\prime}}$ azimuthal zones and sampling $\rm N_t$ times, $\rm N=\rm N_t \times \rm N_r \times \rm {N_{\theta^\prime}}$ computations (numerical integrations) of the ionising flux were needed to calculate the BEL shape evolution due to the Doppler-boost-driven anisotropic illumination of the BLR elements.

\begin{figure}[h!]
    \centering
    \includegraphics[width= \hsize]{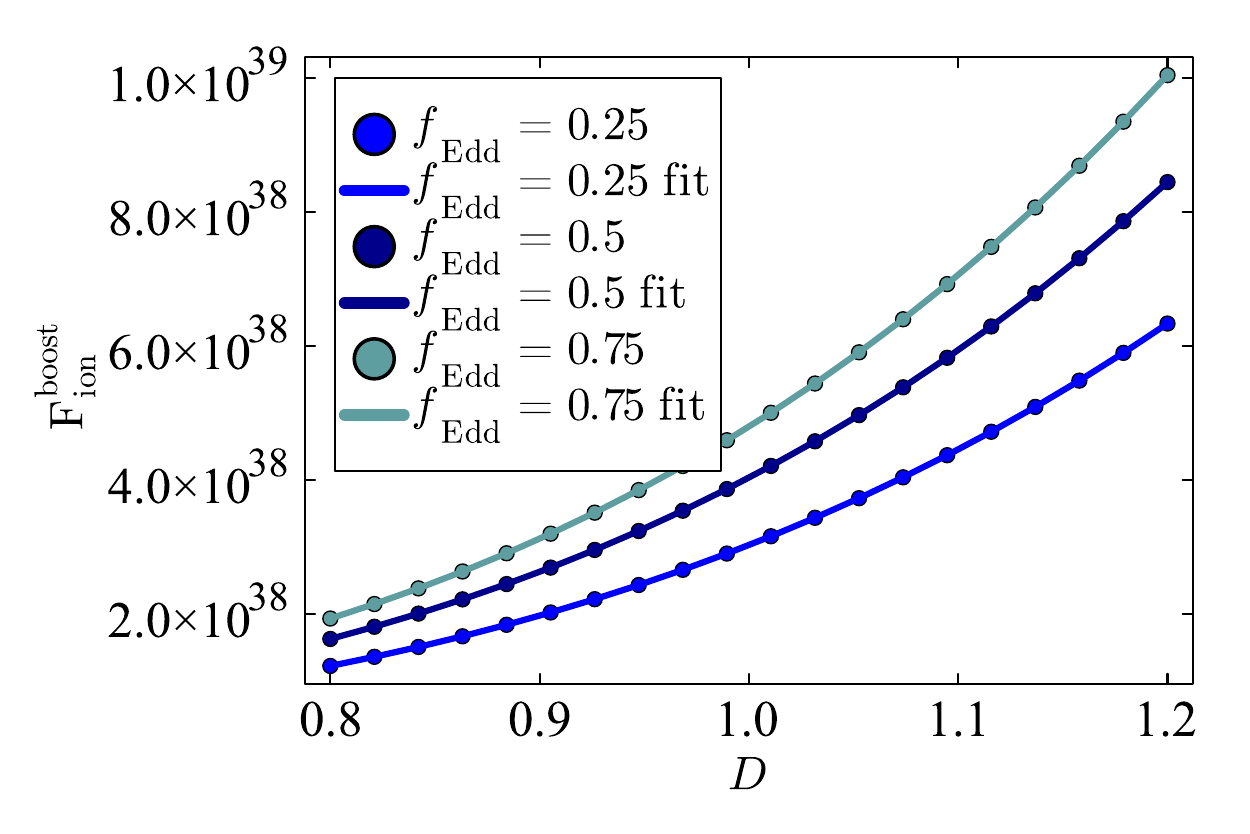}
    \caption{Results for the fit of the relation between the boosted ionising flux, $\rm{F_{ion}^{boost}}$, and the Doppler coefficient for an MBH of $10^6 \rm{M_{\odot}}$ at Eddington ratios equal to $f_{\rm{Edd}}=[0.25,0.50,0.75]$.} 
    \label{fig:fit_doppler}
\end{figure}

Since this operation is computationally expensive, we computed $n \ll N$ integrals for different values of the Doppler coefficient, $D$ (see eq.~ \ref{eq:Doppler_coeff}), and to fit the result using polynomials\footnote{The fit was done using the Julia package Polynomials.jl. The documentation of Polynomials.jl can be found here: \url{https://www.juliapackages.com/p/polynomials}}. In Figure \ref{fig:fit_doppler} we show an example of the results obtained for this fitting procedure for an MBH of $10^6 \  \rm{M_{\odot}}$ at different Eddington ratios. 

 The algorithm proceeded as follows: 
(1) we evaluated $D$ and the area ($\Sigma$) of the element, 
(2) the emissivity, $\epsilon(\xi,\theta)$, of the BLR, and 
(3) the contribution of each element to the observed BEL at each time. This last step was sped up significantly by the fitting procedure for all the models considered. At the same time, step (2) benefited from our fitting too, as the chosen emissivity assumes that $\xi_{\rm{cent}}$ depends on the impinging ionising flux.

The observed flux for the considered BEL from the BLR at a given time, $t$, was computed as the sum of the contributions from each BLR element. We assumed that the contribution from each element is given by a Gaussian in the wavelength domain: 
\begin{equation}
    \rm F_{\rm el}(\lambda) \propto \frac{1}{2\pi\sqrt{\sigma_{\rm line}}} \ e^{-\frac{(\lambda-\lambda_{\rm obs})^2}{2 \sigma_{\rm line}^2}} .
    \label{eq:BEL_gaussian}
\end{equation}
Here, the standard deviation of the Gaussian is given by $\sigma_{\rm{line}}=\sigma \lambda_{\rm{emit}}/c$, where $\sigma$ measures the local velocity dispersion due to small-scale turbulent motions within the BLR disc. In our work, this broadening parameter is given by $\sigma=\upsilon_{\mathrm{Kep}}/10$, where $\upsilon_{\mathrm{Kep}}$ is the Keplerian velocity for that particular BLR element. Finally, $\lambda_{\rm obs}$ is the wavelength of the considered line in the rest frame properly shifted by the motion of the BLR element along the line of sight, $\lambda_{\rm obs}=\lambda_{\rm emit}(1+\upsilon_{\rm los}/c)$\footnote{In principle the Doppler shift effect would turn the originally assumed Gaussian contribution into a slightly broader and asymmetric distribution. However, due to the relatively small Doppler shifts and the small $\sigma/\upsilon_{\mathrm{Kep}}$ ratio assumed, the effect is negligible and it has not been included to speed up the line construction process.}.

In this work, we do not consider general relativistic effects or higher-order special relativistic effects for the BLR elements, as their normalised distance is $\xi \gg 100$ so that $\beta \ll 1$. More specifically, the effects of the transverse redshift and the gravitational redshift of the elements of the BLR disc produce a fractional wavelength shift of the order of $\beta^2$, while the kinematic Doppler shift produces a fractional wavelength shift of the order of $\beta \gg \beta^2$ so that the first two effects can be neglected. Also, we can neglect the effect of light bending for the light rays emitted from the BLR disc as the departure from a straight line is expected to be less than a degree \citep[see][]{Chen89}.

To take care of periodic modulations, we multiplied each Gaussian by the Doppler-boosted ionising flux, $\mathrm{F^{\rm{boost}}_{\rm ion}}$, which strikes the element computed considering the correct retarded time, $t^\prime$ (see Equation~\ref{eq:time_prime}), i.e. the current implementation we assume that each element responds instantaneously and linearly to the continuum flux reaching the element (i.e. we are implicitly assuming that the BLR clouds are optically thick and absorb all the incident ionising radiation). Assuming $\Sigma$ to be the area of an element of the BLR and $\rm{F_{\rm{ion}}}$ to be the observed ionising flux, the contribution from that particular element can be written as
\begin{equation}
    \rm F_{\rm el}(\lambda) = F_{\rm{ion}}^{\rm{boost}}(t^\prime,\xi,\theta^\prime) \ \frac{\epsilon(\xi,\theta^\prime) \Sigma}{2\pi\sqrt{\sigma_{\rm line}}} \ e^{-\frac{(\lambda-\lambda_{\rm obs})^2}{2 \sigma_{\rm line}^2}} ,
    \label{eq:BEL_flux_bin}
\end{equation}
where $\rm{F}_{\rm ion}^{\rm boost}$ is the observed Doppler-boosted ionising flux computed by combining Equations \ref{eq:boosted_integral}, \ref{eq: angle_v_BLR} and \ref{eq:time_prime}.

In the case of a single MBH at the centre of the BLR, we assume that the single MBH emits an intrinsic, periodically varying flux, $F_{\rm{ion}}$, which mimics the ionising continuum light curve produced by the binary. The shape of the contribution from each element is the same as Equation \ref{eq:BEL_flux_bin} with the substitution of $\rm{F}_{\rm{ion}}$ to $\rm{F}^{\rm{boost}}_{\rm{ion}}$.

\subsection{Light curve construction}\label{subsec:light_curves}
The binary signature proposed in this work is based on the delays between the red and blue components of the BEL and the continuum light curve variations. One possible way to compute BEL-component light curves for this purpose is to divide the BEL in half, as is proposed in \cite{Fast_test}. Unfortunately, in the presence of a strong deviation from axisymmetry (such as a strong spiral perturbation) and for the short-period binaries considered in the current analysis, cutting the line in half might result in too large delays between the two BEL sides. As an example, consider the BLR geometry illustrated in Figure \ref{fig:qualitative_cut}, and assume that the region of the BLR with negative abscissa is receding from the observer, while the region with positive abscissa is approaching the observer. In this case, the spiral arm can introduce an arbitrary delay between the ionising continuum and the red light curve, as well as between the ionising continuum and the blue light curve. While the spiral arm perturbation does not affect the velocity field, the location where the bulk of emission is emitted in the receding and approaching sides of the BLR depends on the spiral parameters. 
To minimise the effect of the poorly constrained asymmetry of the BLR, we considered only the wings of the BEL -- or, equivalently, we considered only the elements with a high line-of-sight velocity, highlighted in green in Figure \ref{fig:qualitative_cut}. For these elements, the light-crossing time difference is much smaller than half of a binary period, so that the time delays between the blue and red wings' response will only be due to the phase shifts caused by the binary's Doppler modulation. More specifically, the BEL was cut in the following way: 

\begin{itemize}
    \item Double-peaked lines: the line presents two maxima, one on the blue side (left of the rest-frame wavelength) and one on the red side (right of the rest-frame wavelength). We took the temporal mean of the BEL flux and defined two maxima, 
    $\bar{F}_{\mathrm{blue}}$ and $\bar{F}_{\mathrm{red}}$, occurring at $\lambda_{\mathrm{max,blue}}$ and $\lambda_{\mathrm{max,red}}$, respectively. The BEL was cut at the wavelengths $\lambda_{\mathrm{blue}}$  and $\lambda_{\mathrm{red}}$, where the flux of the BEL is given by $\bar{F}(\lambda_{\mathrm{blue}})=\bar{F}_{\mathrm{blue}}/2$ and  $\bar{F}(\lambda_{\mathrm{red}})=\bar{F}_{\mathrm{red}}/2$, with $\lambda_{\mathrm{blue}}<\lambda_{\mathrm{max,blue}}$ and $\lambda_{\mathrm{red}}>\lambda_{\mathrm{max,red}}$. In Figure \ref{fig:double_peak_cut}, we show an example of how a double-peaked line is cut.
    \begin{figure}[h!]
        \centering
        \includegraphics[width=\hsize]{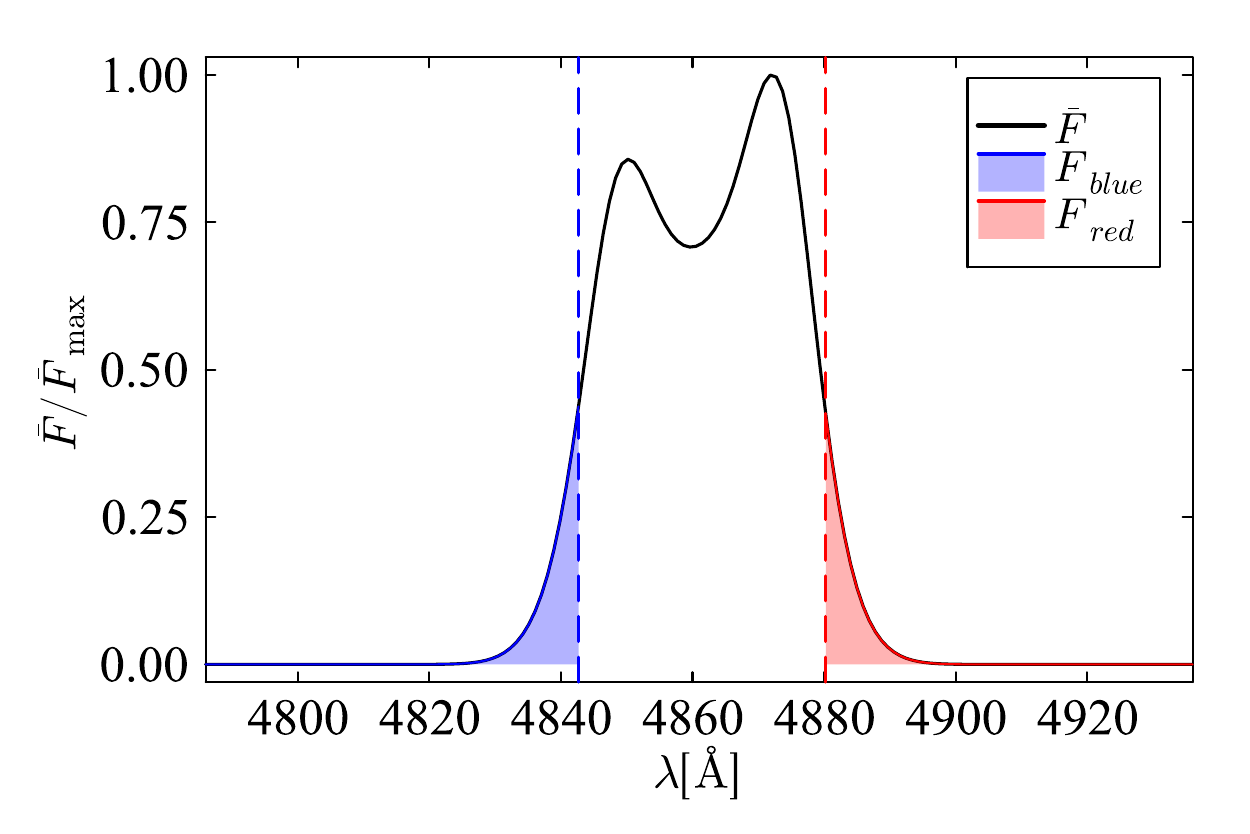}
        \caption{Example of cuts for a double-peaked line. The black line represents the temporal mean of the BEL flux, while the vertical dashed red and blue lines represent the wavelengths at which the line is cut.}
        \label{fig:double_peak_cut}
    \end{figure}
    \item Single-peaked line: the line presents only one maximum. Here, the wavelengths of the two cuts, $\lambda_{\mathrm{blue}}$ and $\lambda_{\mathrm{red}}$, are found by finding the wavelengths where the mean flux is half its maximum value. 
\end{itemize}

The two light curves (‘blue’ and ‘red’) were then computed by integrating the flux over the wavelengths below $\lambda_{\mathrm{blue}}$ and above $\lambda_{\mathrm{red}}$, at each sampled time. Cutting the BEL as described above, the delays between the red and blue light curves are expected to be near zero in the case of a single MBH at the centre of the BLR, as in this scenario the light is emitted isotropically. In the case in which the central source is an MBHB, the emission is beamed along the direction of motion, and the delays between the red and blue light curves are expected to be nearly half the orbital period, independently of the properties of the spiral. This last expectation (confirmed in Section~\ref{sec:results}) depends on the assumption of a coplanar MBHB-BLR system. In the presence of a significant misalignment\footnote{For example, during a transient phase of re-alignment, as is described in \cite{MillerKrolik13}.} the beaming effect onto the BLR would be even clearer, as specific regions of the BLR would be exposed to Doppler-boosted flux only at two specific binary phases, when the MBH velocities point in the direction of the misaligned disc. The frequencies at which such an excess in the reprocessed luminosity would appear would, however, depend on the specific 3D orientation of the observer-binary-circumbinary disc. For circular orbit binaries (as considered in this analysis), there will always be two frequencies on opposite sides of the BEL rest frame centroid, but their exact location (i.e. whether they will be in the core of the lines or in their wings) would depend on the specific configuration.

\section{Results}\label{sec:results}

As we noted in Section \ref{subsec:binary}, in the binary scenario, the emission from the two MBHs is beamed along the direction of motion of the emitter, while in the single black hole scenario, this does not happen (see Figure \ref{fig:qualitative_schemes}). Consider the two innermost elements of the BLR along the line of nodes. If the radiation is beamed and one of the two MBHs emits much more than the other, one element (e.g. the element moving towards the observer) is expected to reprocess the maximum impinging ionising flux after a certain delay given by the BLR light travel time. The receding element is expected to reprocess the maximum ionising continuum half an orbital period after the approaching element. This means that, given the choice of BLR extent assumed here, the response of the red part of the BEL will have a time lag bigger than half a period relative to the ionising continuum light curve. In other words, the peak of the red response is nearer to the next observed peak of the continuum emission (i.e. the boosted contribution of the less luminous MBH), and thus the delay can be seen as negative. 
The lags of the blue and red light curves relative to the ionising continuum will have opposite signs: $\tau_{\mathrm{blue}}>0$ and $\tau_{\mathrm{red}}<0$.
Cutting the line as explained in Section \ref{subsec:light_curves}, we focus on the contribution from the inner part of the BLR near the line of nodes, and thus we expect to see the effect discussed in the example above by measuring the lags between pairs of light curves: the ionising continuum and the red light curve, as well as the ionising continuum and the blue light curve. Depending on the ratio between the BLR size and binary separation, there can be cases in which the delays have the same sign but their difference is nevertheless expected to be around half the binary period.

In the single black hole scenario, the light is not beamed. This means that both the approaching and receding elements will reprocess the impinging ionising flux nearly at the same time.  Although the presence of a spiral arm in the BLR can introduce some differences, causing slight variations in the delays relative to the ionising continuum light curve, these lags will still be comparable and will have the same signs, and, in general, their difference should be negligible. 

The lags between the pair of light curves can be combined in a parameter defined as  
\begin{equation}\label{eq:chi_signature}
    \chi=\frac{|\tau_{\mathrm{red}}-\tau_{\mathrm{blue}}|}{|\tau_{\mathrm{red}}+\tau_{\mathrm{blue}}|} \ .
\end{equation}
This quantity is sensitive to the relative sign between the two time lags, $\tau_{\mathrm{red}}$ and $\tau_{\mathrm{blue}}$. If the signs are discordant, $\chi$ will be larger than $1$, while if the signs are concordant, $\chi$ will be smaller than $1$. In the binary scenario, $\chi$ is expected to spread around 1 and in most cases larger than $1$, depending on the size of the BLR, while in the single MBH scenario, $\chi$ is expected to be around zero, regardless of the size of the BLR. 
This quantity discriminates between binaries and single MBHs with varying intrinsic luminosities:
\begin{equation}\label{eq:signature}
    \begin{split}
        \chi&\gtrapprox 1 \ \Rightarrow \ \mathrm{binary} \\
        \chi& \approx 0 \ \Rightarrow \ \mathrm{single \ MBH}   
    \end{split}
.\end{equation}

\begin{figure}[h!]
    \centering
    \includegraphics[width=\hsize]{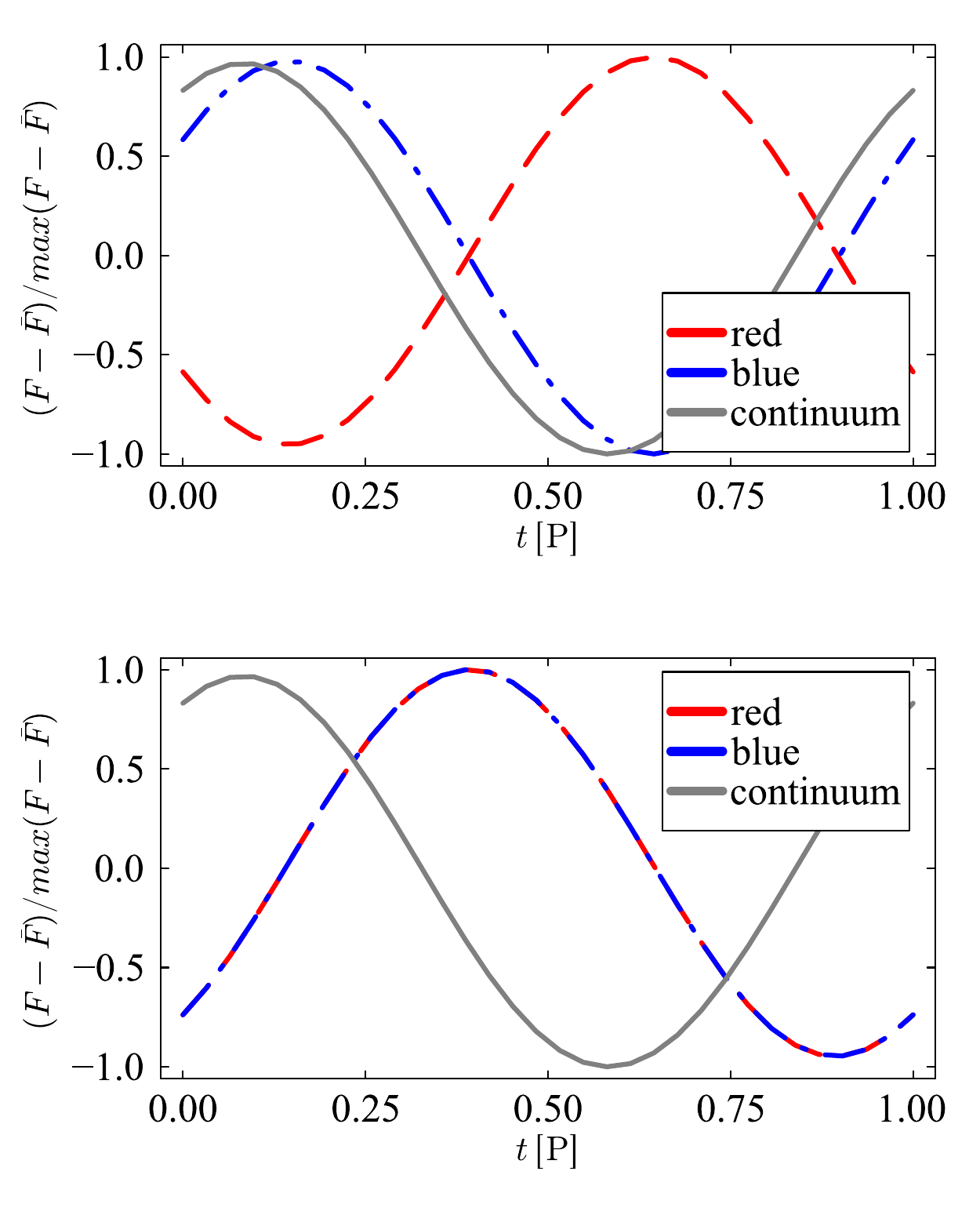}
    \caption{Light curves for the ionising continuum from the central source and for the red and blue BEL wings. The x axis represents time in units of the orbital period, while the y axis represents the normalised flux. Upper panel: Binary scenario with  $M_1=10^7 \ \mathrm{M_{\odot}}$ and $M_2=10^6 \ \mathrm{M_{\odot}}$. The separation is $a=10^{-3.9} \ \mathrm{pc}$. Lower panel: Single MBH scenario with the same ionising continuum as in the binary case.}
    \label{fig:continuuves_binary}
\end{figure}

\begin{figure}[h!]
    \centering
    \includegraphics[width=\hsize]{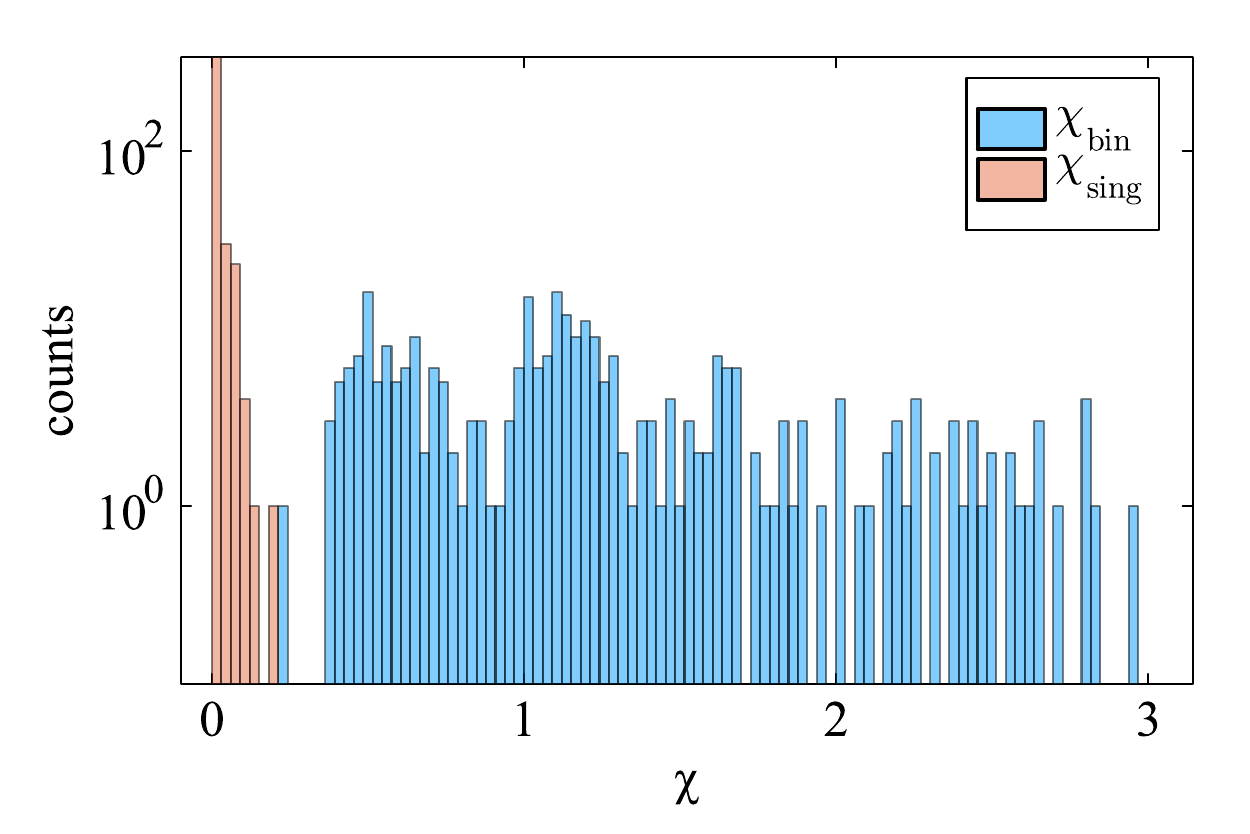}
    \caption{Distribution of the statistic $\chi$ (eq.~\ref{eq:chi_signature}) reflecting the difference in time delays in the blue and red BEL wings, in the single MBH versus the binary scenarios.  }
    \label{fig:Chi_histo_log}
\end{figure}

In Figure~\ref{fig:continuuves_binary} we show examples of light curves in the two cases. The upper panel of Figure~\ref{fig:continuuves_binary} shows the binary scenario (with $M_1=10^7$ M$_\odot$, $q=0.1$ and $f_{\mathrm{Edd,1}}/f_{\mathrm{Edd,2}}\approx 0.02$) where, as expected, the blue light curve responds with a positive delay relative to the continuum light curve, in this case $\chi=1.38$. The lower panel of \ref{fig:continuuves_binary} shows the case of a single MBH with a varying luminosity; in this case, the red and blue light curves respond at nearly the same time, yielding $\chi=0.004$. In Figure \ref{fig:Chi_histo_log} we show the distribution of $\chi$ in the two scenarios, keeping the same MBH masses as in Figure~\ref{fig:continuuves_binary} and changing the BLR characteristic radius and the spiral parameters. We set $R_{\rm{BLR}}$, so that the light travel time varies uniformly between 0.1 and 0.5 times the orbital period, while the spiral parameters vary uniformly in the ranges $A=[0,10]$, $p=[-\pi/2,\pi/2]$, $\delta=[0,2\pi]$, and $\phi_0=[0,2\pi]$ (see Appendix~\ref{app:BLR_emis}).  When considering the binary scenarios of these realisations, we find $\chi<0.5$ in $0.6\%$ of the realisations. In the single MBH scenario, instead $\chi\approx 0$, with $\chi>0.05$ in only $0.07\%$ of the realisations.

The test we propose works for unequal-mass binaries with a mass ratio of $q\lesssim0.3$ but becomes observationally challenging as the mass ratio increases. This is because, to first order, the amplitude of the variability induced by the Doppler boost decreases with an increasing mass ratio, as it results from the sum of sinusoids with similar amplitude but out of phase by half a period.  If the two black holes emit similar spectra with small local spectral curvature (e.g. if the intrinsic emissions have power-law spectral energy distributions), then this results in a near-cancellation of the Doppler modulation. Concerning equal-mass binaries, the test does not work as discussed so far, but can be re-adapted if the BLR shows an axisymmetric emissivity profile. Specifically, a region closer to the bulk of the emission line needs to be considered to construct the red and blue light curves. A study of the effect of mass ratio is shown in Appendix \ref{app:Mass_ratio}.

\begin{figure*}[h!]
    \hspace{0.02\textwidth}
    \subfigure{
        \includegraphics[width=0.45 \textwidth]{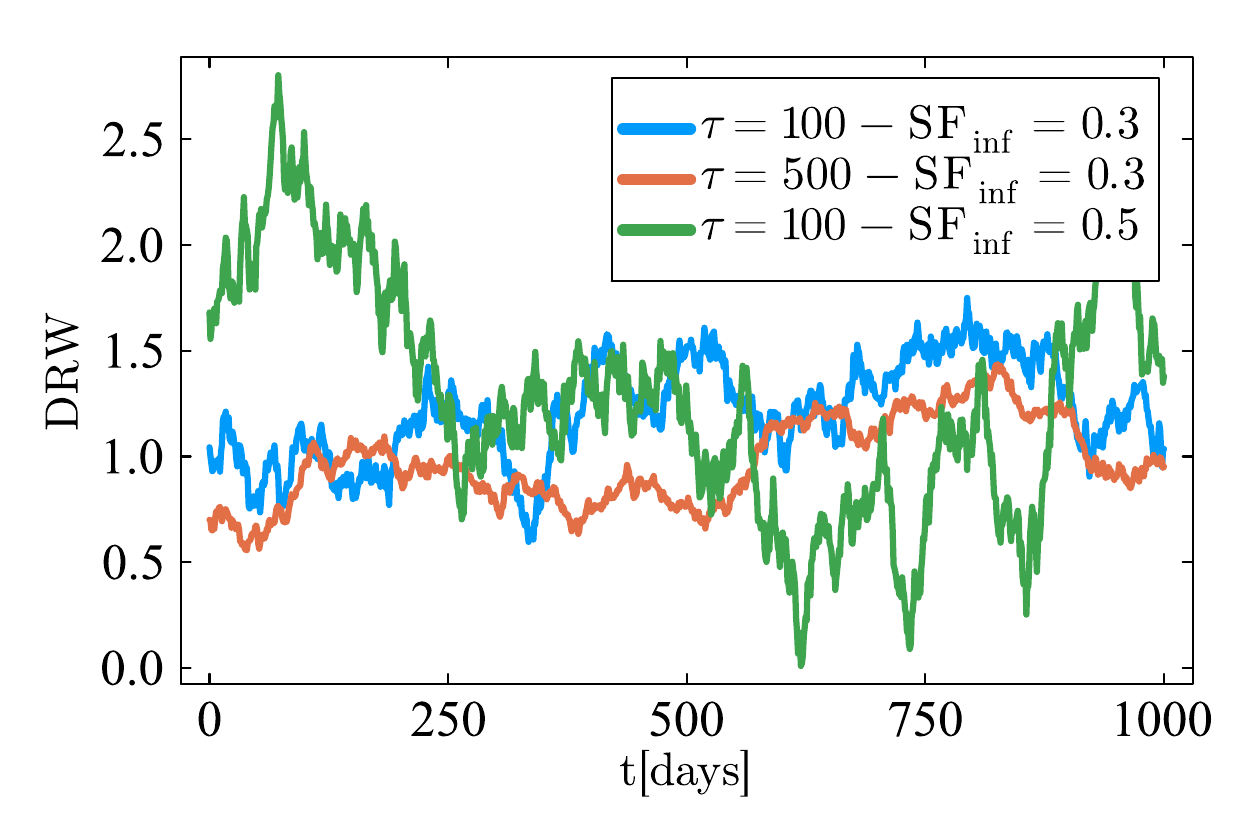}
        }
    \hspace{0.02\textwidth}
    \subfigure{
        \includegraphics[width=0.45 \textwidth]{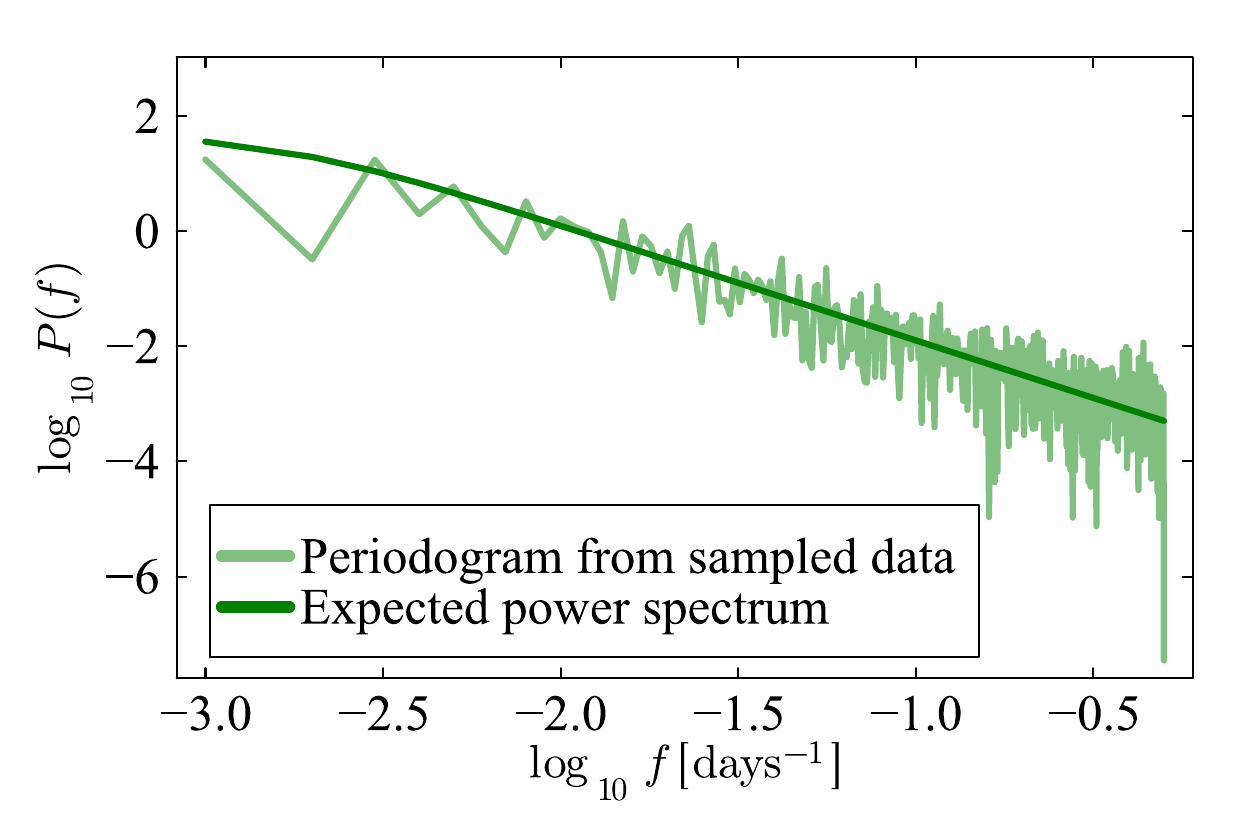}
        }
    \caption{Left: Illustrative examples of DRW time series with different damping timescales and variation amplitudes. Here $\mathrm{SF}_{\inf}$ is the structure function of the process at infinitely large separations in time and is related to the root mean square brightness fluctuation amplitude, $\sigma_{\mathrm{CAR}}$, by $\mathrm{SF}_{\inf}=\sqrt{2}\sigma_{\mathrm{CAR}}$. Right: Expected power spectrum for the DRW superimposed on the periodogram obtained from the green time series of the left image. Note that in both panels, the y axis is reported in arbitrary units.}
    \label{fig:DRW_examples}
\end{figure*}

So far, the effect of intrinsic AGN variability has been neglected. Since we are considering MBH binaries, we expect that, in addition to the variability due to the Doppler boost, the continuum light curve, and consequently the red and blue fluxes from the BLR response, will all be affected by a stochastic red noise in the emission from both MBHs. 
The AGN variability is described well by a red noise, often modelled as a damped random walk \citep[DRW, see][]{DRW}. Since the DRW is an intrinsic property of the continuum source, its effect should be enhanced or diminished by the Doppler boost. The implementation of this effect would ideally enter the Doppler-boosted flux computation. We do not compute the Doppler-boosted flux through a double integration, as it would be too computationally expensive, but through a fitting procedure. Because of this, we cannot implement the DRW before the application of the Doppler boost. To study the effect of the DRW, we implement it more simply as a relative variation in the already Doppler-boosted flux. In Figure \ref{fig:DRW_examples}, we show examples of time series obtained by sampling exclusively from a DRW along with a comparison between the expected power spectrum and the periodogram computed from the sampled DRW time series.
These were used to verify that the chosen sampling of the DRW light curve does not alter the power spectrum of the fluctuations.

\begin{figure}[h!]
    \centering
    \includegraphics[width=0.8 \hsize]{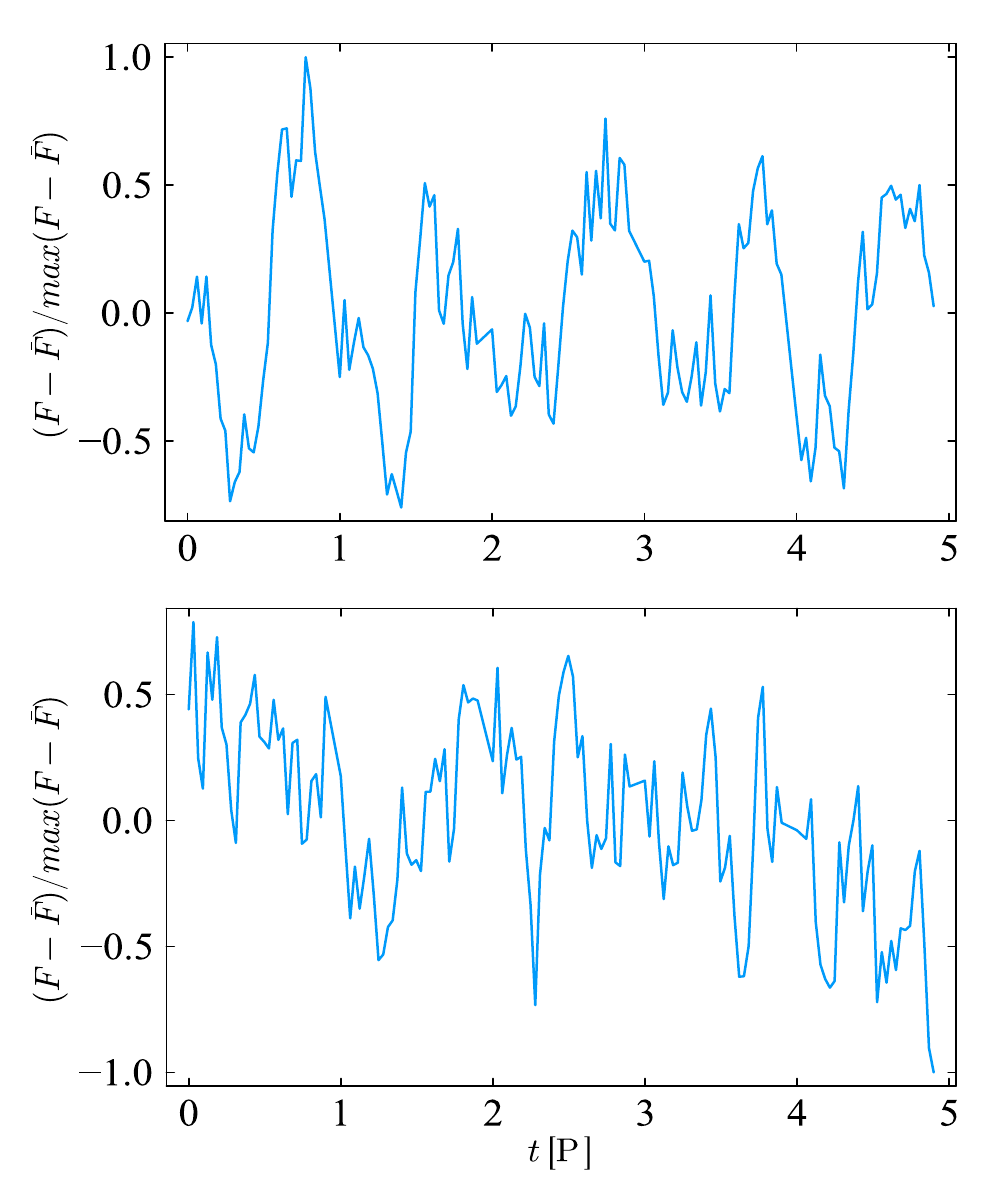}
    \caption{Examples of light curves generated considering both the effects of the Doppler boost and of the variability induced by the damped random walk. In the lower panel, we show the case in which the S/N is 0.28, while in the upper panel, the S/N is 150, and the periodic variability induced by the Doppler boost can be clearly seen.}
    \label{fig:SNRs}
\end{figure}

The parameters we assumed to model the DRW variability are: the mean value of the stochastic process, $\mu=1$, and the characteristic timescale and amplitude of the fluctuation,  $\tau=[100,500] \  \rm{d}$ and $\sigma_{\mathrm{CAR}}=[0.21,0.35]$ for light curves in Figure \ref{fig:DRW_examples}, $\tau=[30,100] \  \rm{d}$ and $\sigma_{\mathrm{CAR}}=[0.03,0.5]$ for light curves in Figure \ref{fig:SNRs}, and finally $\tau=[100,30] \ \rm{d}$ and $\sigma_{\mathrm{CAR}}=[ 0.03,0.01]$ for the MBHs whose light curves are shown in Figure \ref{fig:mediated_fluxes} below. The light curves were sampled using the \href{https://www.astroml.org/modules/generated/astroML.time_series.generate_damped_RW.html}{time\_series.generate\_damped\_RW} function from the \href{https://www.astroml.org}{AstroML} library. We then multiplied the boosted flux from the two MBHs by these values to build the combined light curves, showing good agreement.

The main signature that will identify a binary candidate at separations smaller than $10^{-3} \ \mathrm{pc}$ is the periodic continuum variability.
Red noise introduces two problems. First, it might produce fake periodicities. A distinction between a real periodicity and a fake one can be found by observing the object for a long enough time \citep[see][]{Xin_Haiman}. The second problem is associated with the S/N. If the variability caused by the binary nature of the source is weaker than the variability induced by red noise, it will be difficult to identify the object as a binary candidate using its continuum emission. 
This implies that binaries with separations small enough to have a Doppler boost variability stronger than the noise will have a higher chance of being identified.

In Figure \ref{fig:SNRs} we show two light curves in the case of S/N<1 and S/N>1, where we consider as a signal the average over five orbital periods of the flux obtained by our model ($\bar{f}(t)$), while the noise was computed as the variance of the noisy light curve,
\begin{equation}
    s=\sqrt{\frac{\sum_{i=1}^N(f(t)-\bar{f}(t))^2}{N}},
\end{equation}
where $f(t)$ is the observed noisy flux and $N$ is the number of samples.

Once the period of variation is determined, this information can be used to apply the test proposed in this work to determine whether the source of the light curve is a binary or a single MBH. Specifically, phase-folding the light curve over multiple cycles enhances the periodic features, reduces noise, and allows for a more accurate estimation of the delays between pairs of light curves and their associated uncertainties. Also, by phase-folding the light curve, it is possible to determine whether the periodicity is physical or induced by noise.

In Figure \ref{fig:mediated_fluxes} we show the results of such a procedure for a binary with $q=0.1,\ M=10^7 \ \mathrm{M_{\odot}},\ a=10^{-4.2} \ \mathrm{pc}$ observed for $5,\ 10$, and $20$ periods, with the S/Ns being, respectively, 181, 222, and 242 (a binary with these parameters would merge in about $100 \ \mathrm{yr}$; thus, it would be rarely observed). As can be seen, after phase-folding the light curves become smoother, and uncertainties can be associated with them, allowing for the determination of errors in the measurement of the delays and of the $\chi$ parameter.
\begin{figure*}
    \centering
    \subfigure{
    \includegraphics[width=0.32 \textwidth]{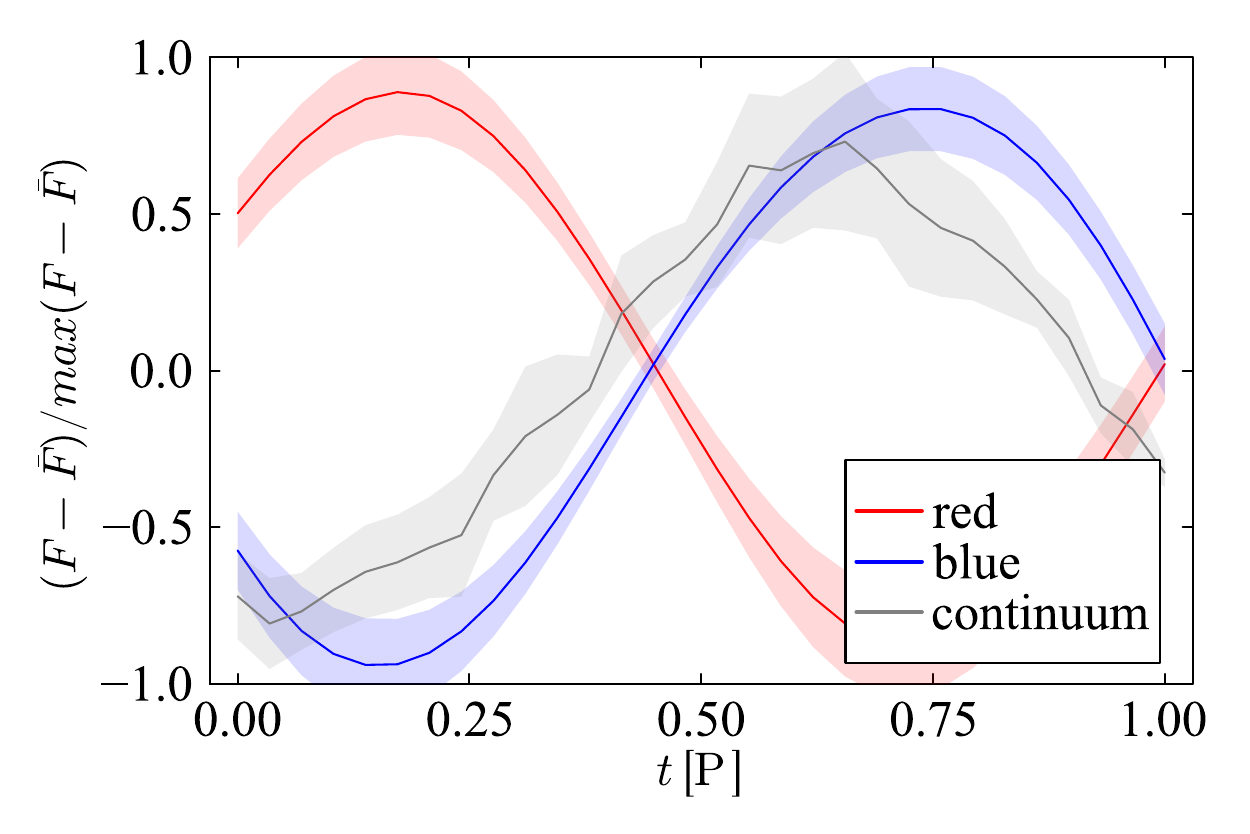}
    }
    \subfigure{
    \includegraphics[width=0.32 \textwidth]{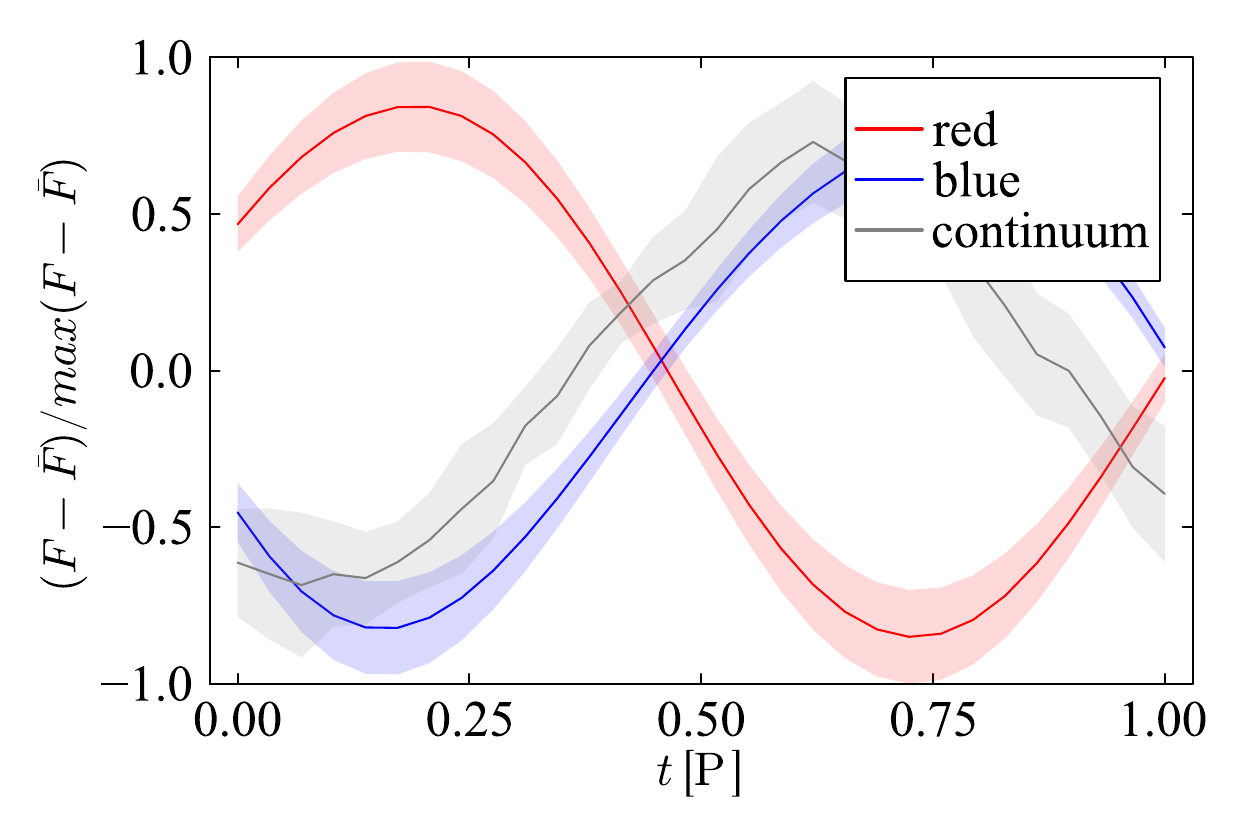}
    }
    \subfigure{
    \includegraphics[width=0.32 \textwidth]{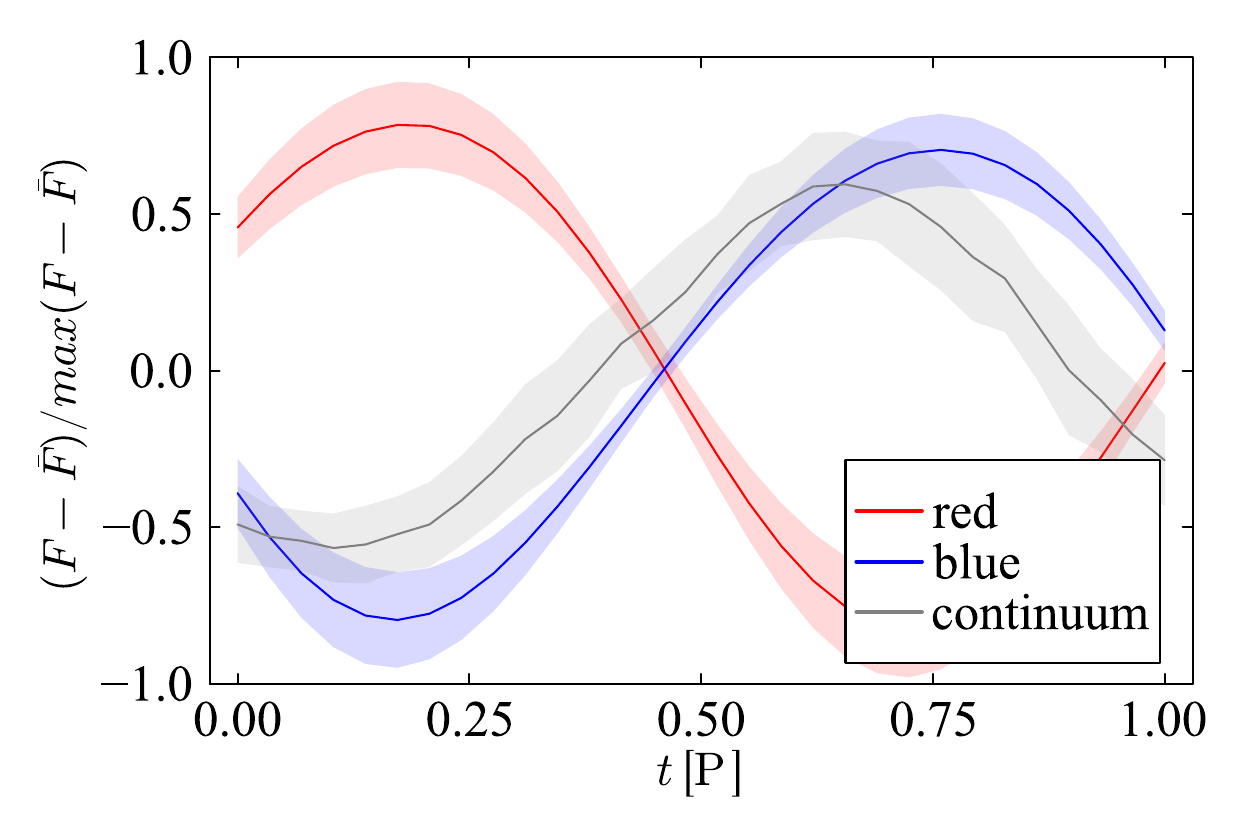}
    }
    
    \caption{Mean ionising continuum, red and blue light curves, and associated uncertainties resulting from the phase-folding procedure observing the source for 5, 10, and 20 periods, respectively (left to right panels), for a binary with $q=0.1,\ M=10^7 \ \mathrm{M_{\odot}},\ a=10^{-4.2} \ \mathrm{pc}$.}
    \label{fig:mediated_fluxes}
\end{figure*}
From the plots, it is clear that the red and blue fluxes show a smoother behaviour compared to the ionising light curve. This is caused by the implementation we are using. As is discussed in Section~\ref{subsec:light_curves}, the red and blue light curves were computed from an emission line built as a sum of the contributions from each element of the BLR. In Section~\ref{subsec:BLR}, we showed that each BLR element responds to the ionising flux emitted at a time that depends on the position of the element. This means that the red and blue fluxes already smooth the noisy continuum, explaining the behaviour shown in the light curves. In this implementation, we did not consider the effect of the S/N due to the measurement noise of such a small part of the spectrum. Taking into account this effect, the uncertainties associated with the red and blue light curves are expected to increase.  

\begin{figure}
    \centering
    \includegraphics[width= \hsize]{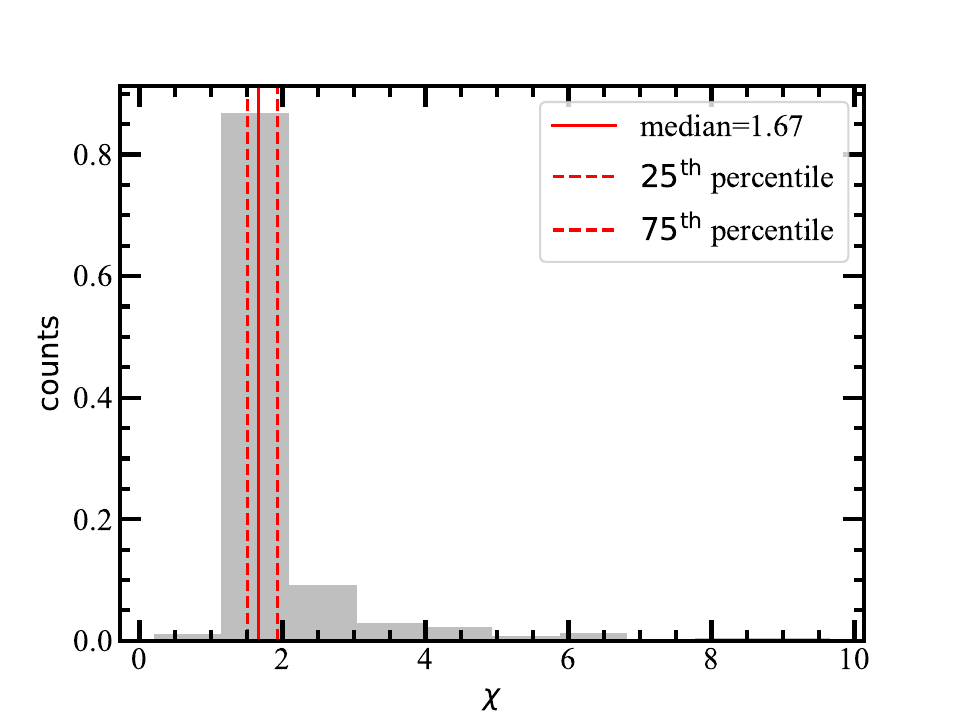}
    \caption{Normalised $\chi$ distribution found through Monte Carlo resampling of fluxes within their observed uncertainties to estimate the median value and uncertainty for a binary observed for 20 periods with $q=0.1,\ M=10^7 \ \mathrm{M_{\odot}},\ a=10^{-4.2} \ \mathrm{pc}$.}
    \label{fig:DRW_histograms}
\end{figure}

As was mentioned in Section~\ref{sec:methodology}, PyCCF can be used to estimate the uncertainties on the time lags between different light curves. Thus, one can use the light curves shown in Figure \ref{fig:mediated_fluxes} to compute the time lags and the uncertainties associated with them, and the $\chi$ parameter to assess whether the source is a binary. In Figure \ref{fig:DRW_histograms} we show the normalised distribution of $\chi$, computed through Monte Carlo simulations, for a noisy light curve observed for 20 periods. Here, the points of the light curves are resampled in a way similar to the one described in \cite{PYCCF}, for the light curves shown in the lower plot of Figure \ref{fig:mediated_fluxes}, showing that the median value of $\chi=1.67$ correctly identifies the source as a binary.

\section{Conclusions and prospects}\label{sec:conclusions}
The goal of this work was to develop an observational signature for MBHBs at orbital separations of the order of $\le 10^{-3} \ \mathrm{pc}$ where a binary is thought to be surrounded by a common BLR. We studied the effect of the Doppler boost caused by the orbital motion of the binary on the BEL evolution. We assumed that the two MBHs are on a circular orbit and that the common BLR has a disc-like structure with a non-axisymmetric emissivity (see Appendix \ref{app:BLR_emis}), where the asymmetry in the BLR is modelled as a spiral arm superimposed on an axisymmetric Gaussian emissivity profile \citep[see][]{sottocorno2025}. The spiral arm can be used to model asymmetric and double-peaked lines \citep{Storchi-Bergmann}, and could, in principle, imprint a different time delay in the different regions of the BEL.

Since the BELs respond to the impinging luminosity emitted by the central engine, information about the nature of the source is expected to be encoded in the shape of the BELs and can be found in reverberation mapping campaigns (with a $\sim$ daily cadence). In this work, we compared the shape evolution of the $\rm{H}\beta$ line in two scenarios: in the first case, the central source is a binary, while in the other, the central source is a single MBH with a varying intrinsic luminosity that mimics the continuum emitted in the binary scenario. We focus on the wings of the BEL, i.e. the regions of the line that are mostly affected by the BLR elements with a high line-of-sight velocity. Doing so reduces the effect of the spiral arm on the emissivity. Integrating the total flux in the selected regions of the BEL, we constructed red and blue light curves and computed the time lags between either of these two time series, relative to the ionising continuum light curve coming directly from the central source.

The main result of this work is that the presence of a binary is encoded in these time lags. Defining the parameter $\chi=|\tau_{red}-\tau_{blue}|/|\tau_{red}+\tau_{blue}|$, one finds that if $\chi \gtrsim 1$, the central object is a MBHB, while if $\chi \approx 0$, the central source is a single MBH. 

After discussing the effectiveness of the test, we introduced stochastic AGN variability, modelled as a damped random walk, to the synthetic light curves. The results show that if the period of variability can be estimated, the light curve can be phase-folded to enhance the periodic features, smooth out the noise, and determine whether the observed periodicity is physical or whether it results from long-term correlations induced by red noise. We show that if the variability is caused by physical mechanisms, the proposed test can correctly identify the source as a binary. 

In Appendix \ref{app:Mass_ratio}, we show how the results change by varying the binary mass ratio. The main conclusion is that the signature we propose works efficiently for unequal mass binaries with a mass ratio of less than $q \sim 0.3$. As $q$ approaches unity, the use of this signature becomes observationally challenging as the sum of the contributions to the continuum from the two MBHs would make the first-order effect of the Doppler boost smaller and smaller by comparison. In the limiting case of an equal mass binary, the continuum is expected to be constant to first order. Thus, the only variability that is shown in Appendix \ref{app:Mass_ratio} is due to second-order effects that are too small to be observed in the presence of noise. Although realistically such effects would be difficult to observe, we showed that the test would still work for unequal mass binaries, while, in the case of an equal mass binary the test can be adapted for an axisymmetric BLR.

The signature proposed in this work must be tested in more general scenarios. First, the Doppler boost is not the only source of variability expected for MBHBs: hydrodynamical simulations show that the accretion rate from a circumbinary disc changes periodically and effects of general relativity, such as gravitational lensing, should be implemented. Second, MBHBs are expected to have some eccentricity, while we considered only circular binaries in this work. Different geometries and emissivity profiles can be used to model the BLR, while more realistic accretion disc models can be implemented, and the effect of the S/N of such a small part of the spectrum should be considered. We plan to test the proposed signature in such scenarios in future work. 

To date, we have not applied the test to real data, as we are not aware of any reverberation mapping campaign on an MBHB candidate selected through a periodically modulated light curve. Multi-epoch spectroscopic campaigns studying AGNs with strong evidence of periodic light curves in current and future time-domain surveys will be the natural test bed for our newly proposed test.

\begin{acknowledgements}
MD acknowledge funding from MIUR under the grant
PRIN 2017-MB8AEZ, and financial support from ICSC – Centro Nazionale di Ricerca in High Performance Computing, Big Data and Quantum Computing, funded by European Union – NextGenerationEU. This study is supported by the Italian Ministry for Research and University (MUR) under Grant 'Progetto Dipartimenti di Eccellenza 2023-2027' (BiCoQ).

FR acknowledges the support from the Next Generation EU funds within the National Recovery and Resilience Plan (PNRR), Mission 4 - Education and Research, Component 2 - From Research to Business (M4C2), Investment Line 3.1 - Strengthening and creation of Research Infrastructures, Project IR0000012 – “CTA+ - Cherenkov Telescope Array Plus.

DJD acknowledges support from the Danish Independent Research Fund through Sapere Aude Starting Grant No. 121587

ZH acknowledges support from NSF grant AST-2006176 and NASA grants 80NSSC22K0822 and 80NSSC24K0440.

\end{acknowledgements}
\bibliographystyle{aa} 
\bibliography{bibliography}

\begin{thebibliography}{59}
\expandafter\ifx\csname natexlab\endcsname\relax\def\natexlab#1{#1}\fi

\bibitem[{{Agazie} {et~al.}(2023){Agazie}, {Anumarlapudi}, {Archibald}, {Arzoumanian}, {Baker}, {B{\'e}csy}, {Blecha}, {Brazier}, {Brook}, {Burke-Spolaor}, {Burnette}, {Case}, {Charisi}, {Chatterjee}, {Chatziioannou}, {Cheeseboro}, {Chen}, {Cohen}, {Cordes}, {Cornish}, {Crawford}, {Cromartie}, {Crowter}, {Cutler}, {Decesar}, {Degan}, {Demorest}, {Deng}, {Dolch}, {Drachler}, {Ellis}, {Ferrara}, {Fiore}, {Fonseca}, {Freedman}, {Garver-Daniels}, {Gentile}, {Gersbach}, {Glaser}, {Good}, {G{\"u}ltekin}, {Hazboun}, {Hourihane}, {Islo}, {Jennings}, {Johnson}, {Jones}, {Kaiser}, {Kaplan}, {Kelley}, {Kerr}, {Key}, {Klein}, {Laal}, {Lam}, {Lamb}, {Lazio}, {Lewandowska}, {Littenberg}, {Liu}, {Lommen}, {Lorimer}, {Luo}, {Lynch}, {Ma}, {Madison}, {Mattson}, {McEwen}, {McKee}, {McLaughlin}, {McMann}, {Meyers}, {Meyers}, {Mingarelli}, {Mitridate}, {Natarajan}, {Ng}, {Nice}, {Ocker}, {Olum}, {Pennucci}, {Perera}, {Petrov}, {Pol}, {Radovan}, {Ransom}, {Ray}, {Romano}, {Sardesai}, {Schmiedekamp}, {Schmiedekamp}, {Schmitz},
  {Schult}, {Shapiro-Albert}, {Siemens}, {Simon}, {Siwek}, {Stairs}, {Stinebring}, {Stovall}, {Sun}, {Susobhanan}, {Swiggum}, {Taylor}, {Taylor}, {Turner}, {Unal}, {Vallisneri}, {van Haasteren}, {Vigeland}, {Wahl}, {Wang}, {Witt}, {Young}, \& {Nanograv Collaboration}}]{Nanograv15}
{Agazie}, G., {Anumarlapudi}, A., {Archibald}, A.~M., {et~al.} 2023, \apjl, 951, L8

\bibitem[{{Amaro-Seoane et al. }(2023)}]{LISA}
{Amaro-Seoane et al. }. 2023, Living Reviews in Relativity, 26, 2

\bibitem[{Antoniadis {et~al.}(2023)Antoniadis, Arumugam, Arumugam, Babak, Bagchi, Bak~Nielsen, Bassa, Bathula, Berthereau, Bonetti, Bortolas, Brook, Burgay, Caballero, Chalumeau, Champion, Chanlaridis, Chen, Cognard, Dandapat, Deb, Desai, Desvignes, Dhanda-Batra, Dwivedi, Falxa, Ferdman, Franchini, Gair, Goncharov, Gopakumar, Graikou, Grießmeier, Guillemot, Guo, Gupta, Hisano, Hu, Iraci, Izquierdo-Villalba, Jang, Jawor, Janssen, Jessner, Joshi, Kareem, Karuppusamy, Keane, Keith, Kharbanda, Kikunaga, Kolhe, Kramer, Krishnakumar, Lackeos, Lee, Liu, Liu, Lyne, McKee, Maan, Main, Mickaliger, Niţu, Nobleson, Paladi, Parthasarathy, Perera, Perrodin, Petiteau, Porayko, Possenti, Prabu, Quelquejay~Leclere, Rana, Samajdar, Sanidas, Sesana, Shaifullah, Singha, Speri, Spiewak, Srivastava, Stappers, Surnis, Susarla, Susobhanan, Takahashi, Tarafdar, Theureau, Tiburzi, van~der Wateren, Vecchio, Venkatraman~Krishnan, Verbiest, Wang, Wang, \& Wu}]{EPTA2023}
Antoniadis, J., Arumugam, P., Arumugam, S., {et~al.} 2023, Astronomy \&amp; Astrophysics, 678, A50

\bibitem[{{Baker} {et~al.}(2019){Baker}, {Haiman}, {Rossi}, {Berger}, {Brandt}, {Breedt}, {Breivik}, {Charisi}, {Derdzinski}, {D'Orazio}, {Ford}, {Greene}, {Hill}, {Holley-Bockelmann}, {Key}, {Kocsis}, {Kupfer}, {Madau}, {Marsh}, {McKernan}, {McWilliams}, {Natarajan}, {Nissanke}, {Noble}, {Phinney}, {Ramsay}, {Schnittman}, {Sesana}, {Shoemaker}, {Stone}, {Toonen}, {Trakhtenbrot}, {Vikhlinin}, \& {Volonteri}}]{Baker-decadal}
{Baker}, J., {Haiman}, Z., {Rossi}, E.~M., {et~al.} 2019, Astro2020 Decadal Survey Sciencey White Paper; BAAS, 51, 123

\bibitem[{{Begelman} {et~al.}(1980){Begelman}, {Blandford}, \& {Rees}}]{Begelman}
{Begelman}, M.~C., {Blandford}, R.~D., \& {Rees}, M.~J. 1980, \nat, 287, 307

\bibitem[{{Bentz} {et~al.}(2009){Bentz}, {Peterson}, {Netzer}, {Pogge}, \& {Vestergaard}}]{Bentz}
{Bentz}, M.~C., {Peterson}, B.~M., {Netzer}, H., {Pogge}, R.~W., \& {Vestergaard}, M. 2009, \apj, 697, 160

\bibitem[{{Capelo} \& {Dotti}(2017)}]{CapeloDotti17}
{Capelo}, P.~R. \& {Dotti}, M. 2017, \mnras, 465, 2643

\bibitem[{{Chen} {et~al.}(1989){Chen}, {Halpern}, \& {Filippenko}}]{Chen89}
{Chen}, K., {Halpern}, J.~P., \& {Filippenko}, A.~V. 1989, \apj, 339, 742

\bibitem[{{De Rosa} {et~al.}(2019){De Rosa}, {Vignali}, {Bogdanovi{\'c}}, {Capelo}, {Charisi}, {Dotti}, {Husemann}, {Lusso}, {Mayer}, {Paragi}, {Runnoe}, {Sesana}, {Steinborn}, {Bianchi}, {Colpi}, {del Valle}, {Frey}, {Gab{\'a}nyi}, {Giustini}, {Guainazzi}, {Haiman}, {Herrera Ruiz}, {Herrero-Illana}, {Iwasawa}, {Komossa}, {Lena}, {Loiseau}, {Perez-Torres}, {Piconcelli}, \& {Volonteri}}]{De_Rosa_dual_AGN}
{De Rosa}, A., {Vignali}, C., {Bogdanovi{\'c}}, T., {et~al.} 2019, \nar, 86, 101525

\bibitem[{{D'Orazio} \& {Charisi}(2023)}]{EM_signatures}
{D'Orazio}, D.~J. \& {Charisi}, M. 2023, arXiv e-prints, arXiv:2310.16896

\bibitem[{{D'Orazio} \& {Duffell}(2021)}]{D'Orazio_Duffell21}
{D'Orazio}, D.~J. \& {Duffell}, P.~C. 2021, \apjl, 914, L21

\bibitem[{{D'Orazio} \& {Haiman}(2017)}]{Dust_lighthouse}
{D'Orazio}, D.~J. \& {Haiman}, Z. 2017, \mnras, 470, 1198

\bibitem[{{D'Orazio} {et~al.}(2015){D'Orazio}, {Haiman}, \& {Schiminovich}}]{Doppler_boost}
{D'Orazio}, D.~J., {Haiman}, Z., \& {Schiminovich}, D. 2015, \nat, 525, 351

\bibitem[{{Dotti} {et~al.}(2023){Dotti}, {Rigamonti}, {Rinaldi}, {Del Pozzo}, {Decarli}, \& {Buscicchio}}]{Fast_test}
{Dotti}, M., {Rigamonti}, F., {Rinaldi}, S., {et~al.} 2023, \aap, 680, A69

\bibitem[{{Duffell} {et~al.}(2020){Duffell}, {D'Orazio}, {Derdzinski}, {Haiman}, {MacFadyen}, {Rosen}, \& {Zrake}}]{m_dot_ratio}
{Duffell}, P.~C., {D'Orazio}, D., {Derdzinski}, A., {et~al.} 2020, \apj, 901, 25

\bibitem[{{Eggleton}(1983)}]{Roche_lobes}
{Eggleton}, P.~P. 1983, \apj, 268, 368

\bibitem[{{Eracleous} {et~al.}(2009){Eracleous}, {Lewis}, \& {Flohic}}]{DPE_Eracleous09}
{Eracleous}, M., {Lewis}, K.~T., \& {Flohic}, H. M.~L.~G. 2009, \nar, 53, 133

\bibitem[{{Haiman} {et~al.}(2023){Haiman}, {Xin}, {Bogdanovi{\'c}}, {Amaro Seoane}, {Bonetti}, {Casey-Clyde}, {Charisi}, {Colpi}, {Davelaar}, {De Rosa}, {D'Orazio}, {Futrowsky}, {Gandhi}, {Graham}, {Greene}, {Habouzit}, {Haggard}, {Holley-Bockelmann}, {Liu}, {Mangiagli}, {Mastrobuono-Battisti}, {McGee}, {Mingarelli}, {Nemmen}, {Palmese}, {Porquet}, {Sesana}, {Stemo}, {Torres-Orjuela}, \& {Zrake}}]{Roman}
{Haiman}, Z., {Xin}, C., {Bogdanovi{\'c}}, T., {et~al.} 2023, arXiv e-prints, arXiv:2306.14990

\bibitem[{{Hutchings} \& {McCall}(1977)}]{Hutchings1977}
{Hutchings}, J.~B. \& {McCall}, M.~L. 1977, \apj, 217, 775

\bibitem[{{Ivezi{\'c}} {et~al.}(2019){Ivezi{\'c}}, {Kahn}, {Tyson}, {Abel}, {Acosta}, {Allsman}, {Alonso}, {AlSayyad}, {Anderson}, {Andrew}, {Angel}, {Angeli}, {Ansari}, {Antilogus}, {Araujo}, {Armstrong}, {Arndt}, {Astier}, {Aubourg}, {Auza}, {Axelrod}, {Bard}, {Barr}, {Barrau}, {Bartlett}, {Bauer}, {Bauman}, {Baumont}, {Bechtol}, {Bechtol}, {Becker}, {Becla}, {Beldica}, {Bellavia}, {Bianco}, {Biswas}, {Blanc}, {Blazek}, {Blandford}, {Bloom}, {Bogart}, {Bond}, {Booth}, {Borgland}, {Borne}, {Bosch}, {Boutigny}, {Brackett}, {Bradshaw}, {Brandt}, {Brown}, {Bullock}, {Burchat}, {Burke}, {Cagnoli}, {Calabrese}, {Callahan}, {Callen}, {Carlin}, {Carlson}, {Chandrasekharan}, {Charles-Emerson}, {Chesley}, {Cheu}, {Chiang}, {Chiang}, {Chirino}, {Chow}, {Ciardi}, {Claver}, {Cohen-Tanugi}, {Cockrum}, {Coles}, {Connolly}, {Cook}, {Cooray}, {Covey}, {Cribbs}, {Cui}, {Cutri}, {Daly}, {Daniel}, {Daruich}, {Daubard}, {Daues}, {Dawson}, {Delgado}, {Dellapenna}, {de Peyster}, {de Val-Borro}, {Digel}, {Doherty}, {Dubois},
  {Dubois-Felsmann}, {Durech}, {Economou}, {Eifler}, {Eracleous}, {Emmons}, {Fausti Neto}, {Ferguson}, {Figueroa}, {Fisher-Levine}, {Focke}, {Foss}, {Frank}, {Freemon}, {Gangler}, {Gawiser}, {Geary}, {Gee}, {Geha}, {Gessner}, {Gibson}, {Gilmore}, {Glanzman}, {Glick}, {Goldina}, {Goldstein}, {Goodenow}, {Graham}, {Gressler}, {Gris}, {Guy}, {Guyonnet}, {Haller}, {Harris}, {Hascall}, {Haupt}, {Hernandez}, {Herrmann}, {Hileman}, {Hoblitt}, {Hodgson}, {Hogan}, {Howard}, {Huang}, {Huffer}, {Ingraham}, {Innes}, {Jacoby}, {Jain}, {Jammes}, {Jee}, {Jenness}, {Jernigan}, {Jevremovi{\'c}}, {Johns}, {Johnson}, {Johnson}, {Jones}, {Juramy-Gilles}, {Juri{\'c}}, {Kalirai}, {Kallivayalil}, {Kalmbach}, {Kantor}, {Karst}, {Kasliwal}, {Kelly}, {Kessler}, {Kinnison}, {Kirkby}, {Knox}, {Kotov}, {Krabbendam}, {Krughoff}, {Kub{\'a}nek}, {Kuczewski}, {Kulkarni}, {Ku}, {Kurita}, {Lage}, {Lambert}, {Lange}, {Langton}, {Le Guillou}, {Levine}, {Liang}, {Lim}, {Lintott}, {Long}, {Lopez}, {Lotz}, {Lupton}, {Lust}, {MacArthur}, {Mahabal},
  {Mandelbaum}, {Markiewicz}, {Marsh}, {Marshall}, {Marshall}, {May}, {McKercher}, {McQueen}, {Meyers}, {Migliore}, {Miller}, {Mills}, {Miraval}, {Moeyens}, {Moolekamp}, {Monet}, {Moniez}, {Monkewitz}, {Montgomery}, {Morrison}, {Mueller}, {Muller}, {Mu{\~n}oz Arancibia}, {Neill}, {Newbry}, {Nief}, {Nomerotski}, {Nordby}, {O'Connor}, {Oliver}, {Olivier}, {Olsen}, {O'Mullane}, {Ortiz}, {Osier}, {Owen}, {Pain}, {Palecek}, {Parejko}, {Parsons}, {Pease}, {Peterson}, {Peterson}, {Petravick}, {Libby Petrick}, {Petry}, {Pierfederici}, {Pietrowicz}, {Pike}, {Pinto}, {Plante}, {Plate}, {Plutchak}, {Price}, {Prouza}, {Radeka}, {Rajagopal}, {Rasmussen}, {Regnault}, {Reil}, {Reiss}, {Reuter}, {Ridgway}, {Riot}, {Ritz}, {Robinson}, {Roby}, {Roodman}, {Rosing}, {Roucelle}, {Rumore}, {Russo}, {Saha}, {Sassolas}, {Schalk}, {Schellart}, {Schindler}, {Schmidt}, {Schneider}, {Schneider}, {Schoening}, {Schumacher}, {Schwamb}, {Sebag}, {Selvy}, {Sembroski}, {Seppala}, {Serio}, {Serrano}, {Shaw}, {Shipsey}, {Sick}, {Silvestri},
  {Slater}, {Smith}, {Smith}, {Sobhani}, {Soldahl}, {Storrie-Lombardi}, {Stover}, {Strauss}, {Street}, {Stubbs}, {Sullivan}, {Sweeney}, {Swinbank}, {Szalay}, {Takacs}, {Tether}, {Thaler}, {Thayer}, {Thomas}, {Thornton}, {Thukral}, {Tice}, {Trilling}, {Turri}, {Van Berg}, {Vanden Berk}, {Vetter}, {Virieux}, {Vucina}, {Wahl}, {Walkowicz}, {Walsh}, {Walter}, {Wang}, {Wang}, {Warner}, {Wiecha}, {Willman}, {Winters}, {Wittman}, {Wolff}, {Wood-Vasey}, {Wu}, {Xin}, {Yoachim}, \& {Zhan}}]{LSST}
{Ivezi{\'c}}, {\v{Z}}., {Kahn}, S.~M., {Tyson}, J.~A., {et~al.} 2019, \apj, 873, 111

\bibitem[{{Kaspi} {et~al.}(2000){Kaspi}, {Smith}, {Netzer}, {Maoz}, {Jannuzi}, \& {Giveon}}]{Kaspi00}
{Kaspi}, S., {Smith}, P.~S., {Netzer}, H., {et~al.} 2000, \apj, 533, 631

\bibitem[{{Kelley} {et~al.}(2021){Kelley}, {D'Orazio}, \& {Di Stefano}}]{Self_lensing}
{Kelley}, L.~Z., {D'Orazio}, D.~J., \& {Di Stefano}, R. 2021, \mnras, 508, 2524

\bibitem[{Kelly {et~al.}(2009)Kelly, Bechtold, \& Siemiginowska}]{DRW}
Kelly, B.~C., Bechtold, J., \& Siemiginowska, A. 2009, The Astrophysical Journal, 698, 895

\bibitem[{{MacFadyen} \& {Milosavljevi{\'c}}(2008)}]{Accretion_variability}
{MacFadyen}, A.~I. \& {Milosavljevi{\'c}}, M. 2008, \apj, 672, 83

\bibitem[{Maggiore(2008)}]{maggiore_GW}
Maggiore, M. 2008, Gravitational Waves: Volume 1: Theory and Experiments, Gravitational Waves (OUP Oxford)

\bibitem[{{McLure} \& {Dunlop}(2001)}]{McLure01}
{McLure}, R.~J. \& {Dunlop}, J.~S. 2001, \mnras, 327, 199

\bibitem[{{Miller} \& {Krolik}(2013)}]{MillerKrolik13}
{Miller}, M.~C. \& {Krolik}, J.~H. 2013, \apj, 774, 43

\bibitem[{{Moloney} \& {Shull}(2014)}]{Moloney2014}
{Moloney}, J. \& {Shull}, J.~M. 2014, \apj, 793, 100

\bibitem[{{Mu{\~n}oz} {et~al.}(2019){Mu{\~n}oz}, {Miranda}, \& {Lai}}]{Munoz19}
{Mu{\~n}oz}, D.~J., {Miranda}, R., \& {Lai}, D. 2019, \apj, 871, 84

\bibitem[{{Nguyen} \& {Bogdanovi{\'c}}(2016)}]{Nguyen-Bogdanovic_1}
{Nguyen}, K. \& {Bogdanovi{\'c}}, T. 2016, \apj, 828, 68

\bibitem[{{Nguyen} {et~al.}(2019){Nguyen}, {Bogdanovi{\'c}}, {Runnoe}, {Eracleous}, {Sigurdsson}, \& {Boroson}}]{Nguyen-Bogdanovic_2}
{Nguyen}, K., {Bogdanovi{\'c}}, T., {Runnoe}, J.~C., {et~al.} 2019, \apj, 870, 16

\bibitem[{{Novikov} \& {Thorne}(1973)}]{Novikov_Thorne}
{Novikov}, I.~D. \& {Thorne}, K.~S. 1973, in Black Holes (Les Astres Occlus), 343--450

\bibitem[{{Page} \& {Thorne}(1974)}]{Page1974}
{Page}, D.~N. \& {Thorne}, K.~S. 1974, \apj, 191, 499

\bibitem[{{Patterson} {et~al.}(1993){Patterson}, {Halpern}, \& {Shambrook}}]{Patterson93}
{Patterson}, J., {Halpern}, J., \& {Shambrook}, A. 1993, \apj, 419, 803

\bibitem[{{Peterson} {et~al.}(1998){Peterson}, {Wanders}, {Horne}, {Collier}, {Alexander}, {Kaspi}, \& {Maoz}}]{PYCCF}
{Peterson}, B.~M., {Wanders}, I., {Horne}, K., {et~al.} 1998, \pasp, 110, 660

\bibitem[{{Rigamonti} {et~al.}(2025){Rigamonti}, {Severgnini, Paola}, {Sottocorno, Erika}, {Dotti, Massimo}, {Covino, Stefano}, {Landoni, Marco}, {Bertassi, Lorenzo}, {Braito, Valentina}, {Cicone, Claudia}, {Cupani, Guido}, {De Rosa, Alessandra}, {Della Ceca, Roberto}, {Ighina, Luca}, {Singh, Jasbir}, \& {Vignali, Cristian}}]{Rigamonti2025}
{Rigamonti}, F., {Severgnini, Paola}, {Sottocorno, Erika}, {et~al.} 2025, A\&A, 693, A117

\bibitem[{{Robinson}(1976)}]{Robinson1976}
{Robinson}, E.~L. 1976, \araa, 14, 119

\bibitem[{{Rodriguez} {et~al.}(2006){Rodriguez}, {Taylor}, {Zavala}, {Peck}, {Pollack}, \& {Romani}}]{Rodriguez_2006}
{Rodriguez}, C., {Taylor}, G.~B., {Zavala}, R.~T., {et~al.} 2006, \apj, 646, 49

\bibitem[{{Roedig} {et~al.}(2011){Roedig}, {Dotti}, {Sesana}, {Cuadra}, \& {Colpi}}]{Roedig11}
{Roedig}, C., {Dotti}, M., {Sesana}, A., {Cuadra}, J., \& {Colpi}, M. 2011, \mnras, 415, 3033

\bibitem[{{Runnoe} {et~al.}(2013){Runnoe}, {Brotherton}, {Shang}, {Wills}, \& {DiPompeo}}]{Runnoe13}
{Runnoe}, J.~C., {Brotherton}, M.~S., {Shang}, Z., {Wills}, B.~J., \& {DiPompeo}, M.~A. 2013, \mnras, 429, 135

\bibitem[{{Runnoe} {et~al.}(2015){Runnoe}, {Eracleous}, {Mathes}, {Pennell}, {Boroson}, {Sigur{\dh}sson}, {Bogdanovi{\'c}}, {Halpern}, \& {Liu}}]{Truncation_radius}
{Runnoe}, J.~C., {Eracleous}, M., {Mathes}, G., {et~al.} 2015, \apjs, 221, 7

\bibitem[{{Sandrinelli} {et~al.}(2016){Sandrinelli}, {Covino}, {Dotti}, \& {Treves}}]{Sandrinelli16}
{Sandrinelli}, A., {Covino}, S., {Dotti}, M., \& {Treves}, A. 2016, \aj, 151, 54

\bibitem[{{Shankar} {et~al.}(2013){Shankar}, {Weinberg}, \& {Miralda-Escud{\'e}}}]{Shankar2013}
{Shankar}, F., {Weinberg}, D.~H., \& {Miralda-Escud{\'e}}, J. 2013, \mnras, 428, 421

\bibitem[{{Siwek} {et~al.}(2023){Siwek}, {Weinberger}, \& {Hernquist}}]{Siwek}
{Siwek}, M., {Weinberger}, R., \& {Hernquist}, L. 2023, \mnras, 522, 2707

\bibitem[{Sottocorno {et~al.}(2025)Sottocorno, Ogborn, Bertassi, Rigamonti, Eracleous, \& Dotti}]{sottocorno2025}
Sottocorno, E., Ogborn, M., Bertassi, L., {et~al.} 2025, A\&A [\eprint[arXiv]{2504.06340}], submitted

\bibitem[{{Steeghs} {et~al.}(1997){Steeghs}, {Harlaftis}, \& {Horne}}]{steeghs97}
{Steeghs}, D., {Harlaftis}, E.~T., \& {Horne}, K. 1997, \mnras, 290, L28

\bibitem[{{Storchi-Bergmann} {et~al.}(2003){Storchi-Bergmann}, {Nemmen da Silva}, {Eracleous}, {Halpern}, {Wilson}, {Filippenko}, {Ruiz}, {Smith}, \& {Nagar}}]{Storchi-Bergmann}
{Storchi-Bergmann}, T., {Nemmen da Silva}, R., {Eracleous}, M., {et~al.} 2003, \apj, 598, 956

\bibitem[{{Storchi-Bergmann} {et~al.}(2017){Storchi-Bergmann}, {Schimoia}, {Peterson}, {Elvis}, {Denney}, {Eracleous}, \& {Nemmen}}]{Disc_BLR}
{Storchi-Bergmann}, T., {Schimoia}, J.~S., {Peterson}, B.~M., {et~al.} 2017, \apj, 835, 236

\bibitem[{{Storey} \& {Hummer}(1995)}]{Storey1995}
{Storey}, P.~J. \& {Hummer}, D.~G. 1995, \mnras, 272, 41

\bibitem[{{Sun} {et~al.}(2018){Sun}, {Grier}, \& {Peterson}}]{pythoncode18}
{Sun}, M., {Grier}, C.~J., \& {Peterson}, B.~M. 2018, {PyCCF: Python Cross Correlation Function for reverberation mapping studies}, Astrophysics Source Code Library, record ascl:1805.032

\bibitem[{{Trindade Falc{\~a}o} {et~al.}(2024){Trindade Falc{\~a}o}, {Turner}, {Kraemer}, {Reeves}, {Braito}, {Schmitt}, \& {Feuillet}}]{Trindade24}
{Trindade Falc{\~a}o}, A., {Turner}, T.~J., {Kraemer}, S.~B., {et~al.} 2024, \apj, 972, 185

\bibitem[{Vaughan {et~al.}(2016)Vaughan, Uttley, Markowitz, Huppenkothen, Middleton, Alston, Scargle, \& Farr}]{False_periodicies}
Vaughan, S., Uttley, P., Markowitz, A.~G., {et~al.} 2016, Monthly Notices of the Royal Astronomical Society, 461, 3145

\bibitem[{{Verbiest} {et~al.}(2016){Verbiest}, {Lentati}, {Hobbs}, {van Haasteren}, {Demorest}, {Janssen}, {Wang}, {Desvignes}, {Caballero}, {Keith}, {Champion}, {Arzoumanian}, {Babak}, {Bassa}, {Bhat}, {Brazier}, {Brem}, {Burgay}, {Burke-Spolaor}, {Chamberlin}, {Chatterjee}, {Christy}, {Cognard}, {Cordes}, {Dai}, {Dolch}, {Ellis}, {Ferdman}, {Fonseca}, {Gair}, {Garver-Daniels}, {Gentile}, {Gonzalez}, {Graikou}, {Guillemot}, {Hessels}, {Jones}, {Karuppusamy}, {Kerr}, {Kramer}, {Lam}, {Lasky}, {Lassus}, {Lazarus}, {Lazio}, {Lee}, {Levin}, {Liu}, {Lynch}, {Lyne}, {Mckee}, {McLaughlin}, {McWilliams}, {Madison}, {Manchester}, {Mingarelli}, {Nice}, {Os{\l}owski}, {Palliyaguru}, {Pennucci}, {Perera}, {Perrodin}, {Possenti}, {Petiteau}, {Ransom}, {Reardon}, {Rosado}, {Sanidas}, {Sesana}, {Shaifullah}, {Shannon}, {Siemens}, {Simon}, {Smits}, {Spiewak}, {Stairs}, {Stappers}, {Stinebring}, {Stovall}, {Swiggum}, {Taylor}, {Theureau}, {Tiburzi}, {Toomey}, {Vallisneri}, {van Straten}, {Vecchio}, {Wang}, {Wen}, {You},
  {Zhu}, \& {Zhu}}]{Verbiest2016}
{Verbiest}, J.~P.~W., {Lentati}, L., {Hobbs}, G., {et~al.} 2016, \mnras, 458, 1267

\bibitem[{Ward {et~al.}(2024)Ward, Gezari, Nugent, Kerr, Eracleous, Frederick, Hammerstein, Graham, van Velzen, Kasliwal, Laher, Masci, Purdum, Racine, \& Smith}]{Ward_2024}
Ward, C., Gezari, S., Nugent, P., {et~al.} 2024, The Astrophysical Journal, 961, 172

\bibitem[{Ward {et~al.}(2025)Ward, Koss, Eracleous, Trakhtenbrot, Bauer, Caglar, Harrison, Jana, Kakkad, Magno, del Moral-Castro, Mushotzky, Oh, Peca, Powell, Ricci, Rojas, Smith, Stern, Treister, \& Urry}]{ward2025}
Ward, C., Koss, M.~J., Eracleous, M., {et~al.} 2025, BASS LII: The prevalence of double-peaked broad lines at low accretion rates among hard X-ray selected AGN

\bibitem[{{Wills} \& {Browne}(1986)}]{Wills86}
{Wills}, B.~J. \& {Browne}, I.~W.~A. 1986, \apj, 302, 56

\bibitem[{{Xin} \& {Haiman}(2021)}]{Xin_forecast}
{Xin}, C. \& {Haiman}, Z. 2021, \mnras, 506, 2408

\bibitem[{{Xin} \& {Haiman}(2024)}]{Xin_Haiman}
{Xin}, C. \& {Haiman}, Z. 2024, \mnras, 533, 3164

\bibitem[{{Zrake} {et~al.}(2021){Zrake}, {Tiede}, {MacFadyen}, \& {Haiman}}]{Zrake21}
{Zrake}, J., {Tiede}, C., {MacFadyen}, A., \& {Haiman}, Z. 2021, \apjl, 909, L13

\end{thebibliography}

\begin{appendix}

\section{BLR emissivity profile} \label{app:BLR_emis}
The BLR emission is modelled using the modification of the prescription of \cite{Storchi-Bergmann} proposed by \cite{sottocorno2025} (see also \citeauthor{Rigamonti2025} \citeyear{Rigamonti2025} for an application of the model to real data).
In this model, the BLR emissivity features a spiral pattern superimposed on an otherwise axisymmetric disc-like BLR. More specifically, the assumed emissivity profile is given by
\begin{equation}
\begin{split}
\epsilon(\xi,\theta) = &\ \frac{1}{\sqrt{2\pi}\sigma_{\mathrm{cent}}} \frac{1}{\xi} 
\ \exp{\left[-\frac{(\xi-\xi_{\rm cent})^2}{2\sigma_{\rm cent}^2}\right]} \left\{  1 + \frac{A}{2} \exp^{\left[-\frac{4 \ln 2}{\delta^2}(\phi-\psi_0)^2\right]} \right. \\
& \left. + \frac{A}{2} \exp{\left[-\frac{4 \ln 2}{\delta^2}(2\pi -\phi+\psi_0)^2\right]} \right\},
\end{split}
\label{eq:BLR_emissivity}
\end{equation}
where $\xi$ is the radial distance from either the single MBH or the MBHB centre of mass in units of gravitational radii\footnote{The total mass of the MBHB (M) is used for the rescaling in the binary scenario as $\xi=R c^2 / G M$.}, while $\phi$ is the azimuthal co-ordinate in the BLR plane. Here, the spiral perturbation is intended as a proxy of a broad family of perturbations that make some portion of the disc brighter than the rest of the disc. Spirals are however particularly appealing, as their development in discs are observed in a variety of systems, from spiral galaxies to discs around cataclysmic variables \citep[e.g.][]{Patterson93, steeghs97}, and their presence allows for the description of asymmetric broad lines as seen in observations \citep[e.g.][and references therein]{Disc_BLR, Ward_2024, ward2025}. The axisymmetric term of the profile is given by a Gaussian, centred at the normalised radius $\xi_{\rm cent}$ assumed to be equal to the characteristic radius of the BLR ($\xi_{\rm cent}=R_{\rm{BLR}}$), divided by the normalised distance. Using this profile, we automatically obtain a characteristic radius of the BLR consistent with the empirically inferred luminosity-radius relation \citep{Bentz}, once the relation between $\xi_{\rm cent}$ and ionising source luminosity is specified \citep[see][]{sottocorno2025}.

\begin{figure}[h!]
    \centering
    \includegraphics[width=\hsize]{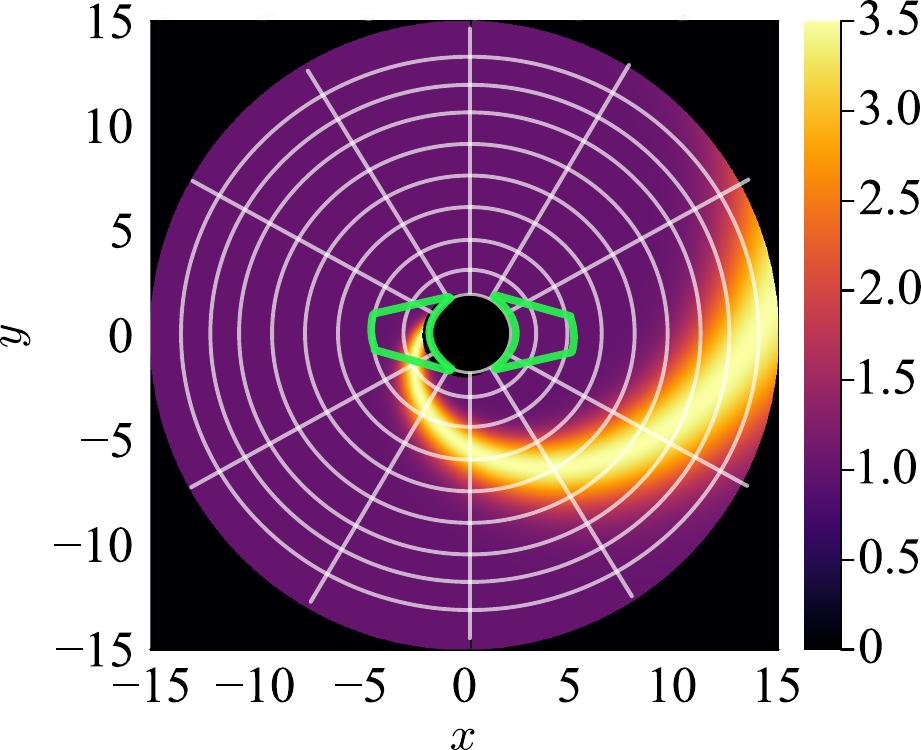}
    \caption{Example of a BLR with a spiral arm. The emissivity is reported in units of its minimum value (on a linear colour gradient) and is set to $0$ in regions where the BLR is not present. The grid elements responsible for the considered red and blue parts of the BEL are highlighted in green (see Section \ref{subsec:light_curves}).}
    \label{fig:qualitative_cut}
\end{figure}

The spiral arm is parametrised by its azimuthal width (or FWHM, $\delta$) and an azimuthal reference co-ordinate $\psi_0(\xi)$. Here $\psi_0=\phi_0+\log(\xi/\xi_{\mathrm{sp}})/\tan(p)$, where$\phi_0$ is the azimuthal position of the spiral at its innermost normalised radius $\xi_{\mathrm{sp}}$ and $p$ is the spiral pitch angle. In this work, we assume $\xi_{\mathrm{sp}}$ to coincide with the innermost radius of the BLR. Finally, the parameter $A$ represents the brightness contrast between the spiral arm and the underlying, axisymmetric disc, so that in total the spiral shape of the BLR is specified by five parameters. In Figure \ref{fig:qualitative_cut}, we show an example of the emissivity profile with a spiral pattern for the set of parameters reported in Table \ref{tab:spirlal_params}.
\begin{table}[h!]
    \centering
    \caption{Spiral parameters used to generate the BLR reported in Figure \ref{fig:qualitative_cut}.}
    \begin{tabular}{|c|c|c|c|c|}
        \hline
        $A$ & $\delta$ & $p$ & $\phi_0$ & $\xi_{sp}$\\
        \hline
        5 & 40° & 15° & 90° & $R_{\rm{BLR}}$\\
        \hline
    \end{tabular}
    \label{tab:spirlal_params}
\end{table}

\section{Effect of the mass ratio}\label{app:Mass_ratio}
In our model, the binary mass ratio determines the accretion rate of the two MBHs. The Eddington ratios at which the two MBHs are emitting are fixed by equation \ref{eq:Edd_ratio_frac}. As the mass ratio approaches unity, the emission from the two MBHs is therefore more and more similar.
\ Hence, as $q$ increases, the emission from the MBH with the lower accretion rate will become more important. Since the emission from the two MBHs will vary sinusoidally with the same frequency, the orbital one, but with a phase difference of $\pi$, the observed light curves will have smaller and smaller amplitudes.
If the amplitude becomes small enough, second-order effects will become visible. In the context of Doppler boosting for a binary, the second-order effect is a sinusoid with a periodicity equal to half the orbital period. Thus, when summing the contributions from the two MBHs, both the continuum and BEL light curves will not necessarily look like sinusoids with a period equal to the orbital period of the binary anymore. 
\begin{figure*}
    \centering
    \subfigure[Binary with $q=0.25$]{
    \includegraphics[width= 7 cm]{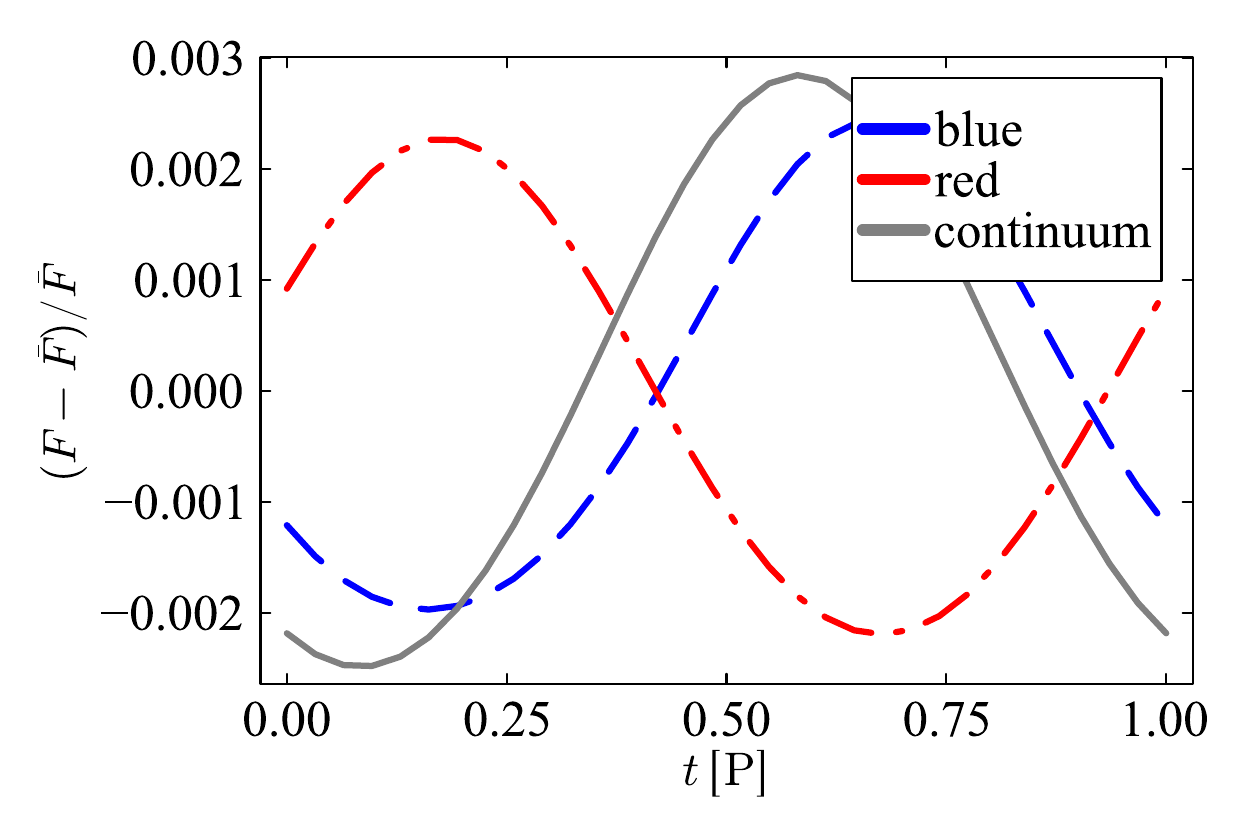}
    }
    \subfigure[Single MBH with the same continuum light curve as the binary case where $q=0.25$]{
    \includegraphics[width=7 cm]{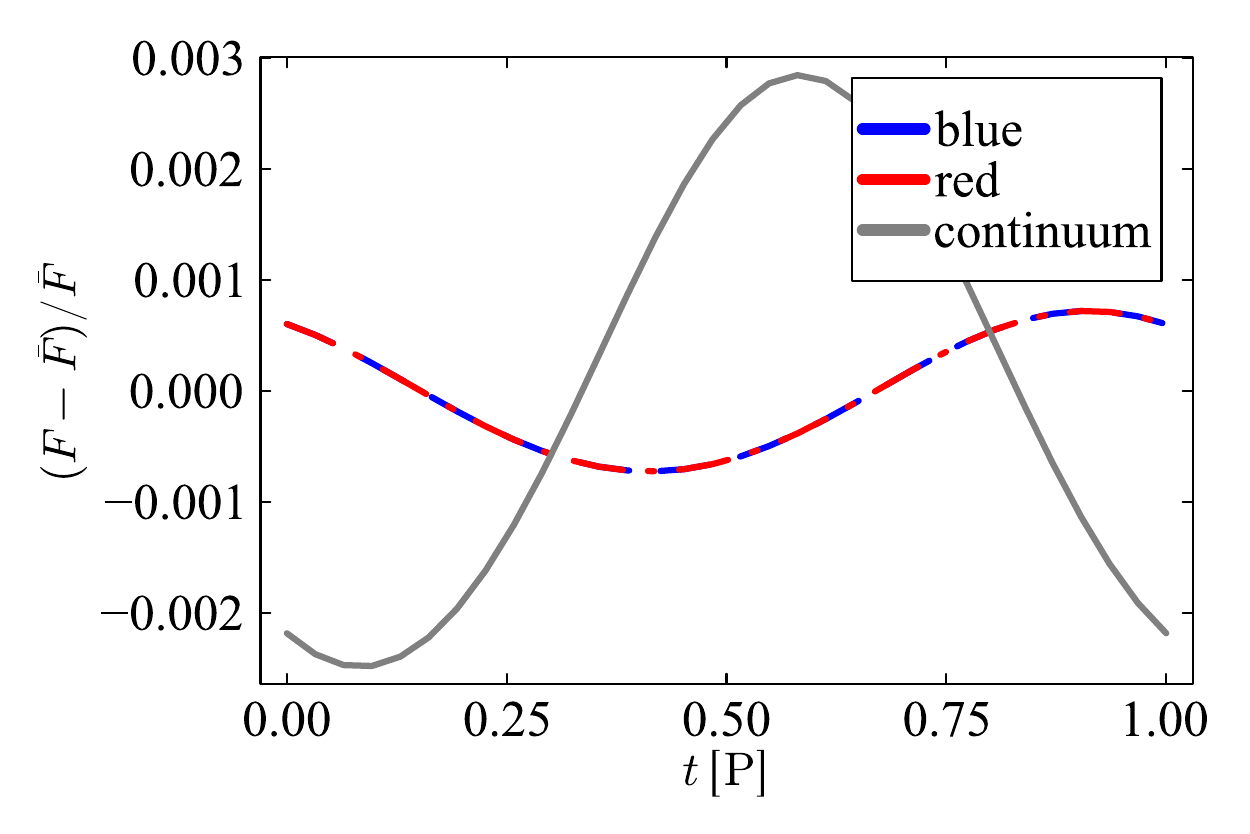}
    }
    \subfigure[Binary with $q=0.55$]{
    \includegraphics[width=7 cm]{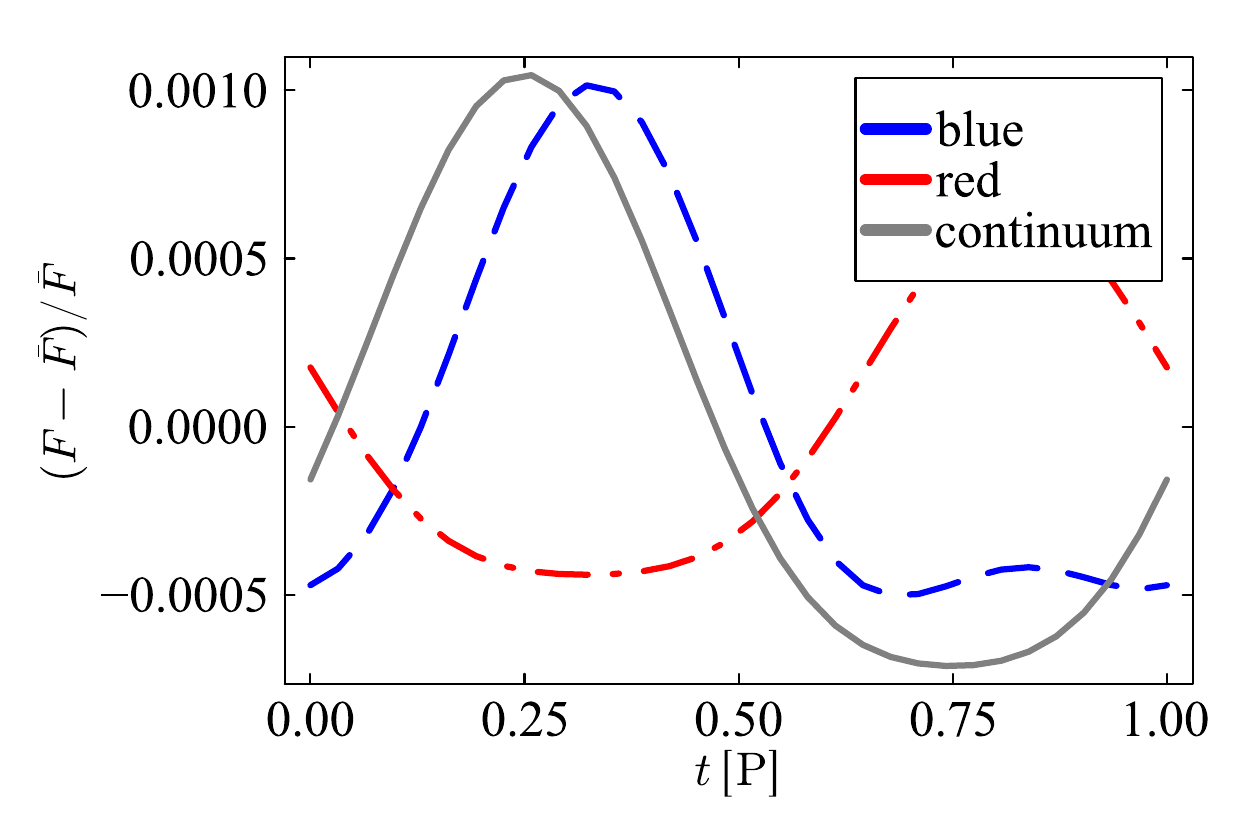}
    }
    \subfigure[Single MBH with the same continuum light curve as the binary case where $q=0.55$]{
    \includegraphics[width=7 cm]{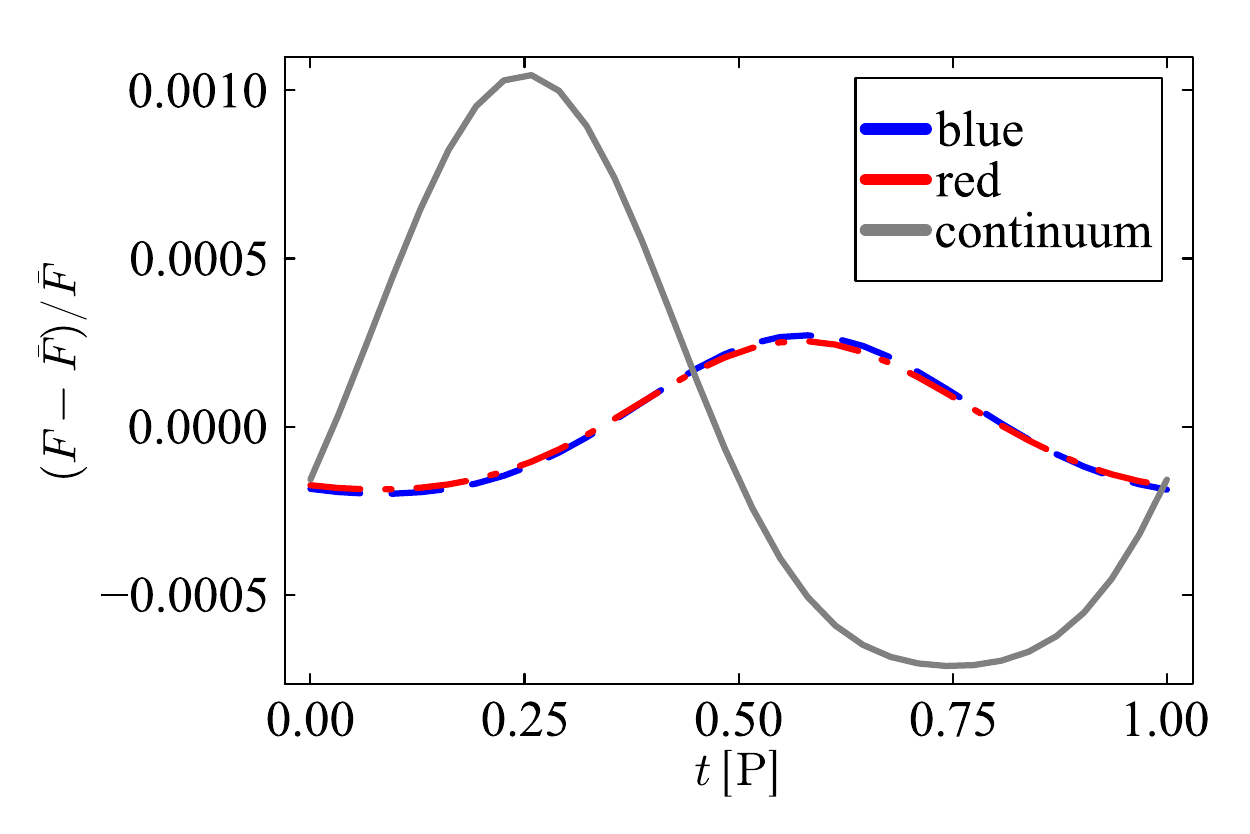}
    }
    \subfigure[Binary with $q=0.75$]{
    \includegraphics[width=7 cm]{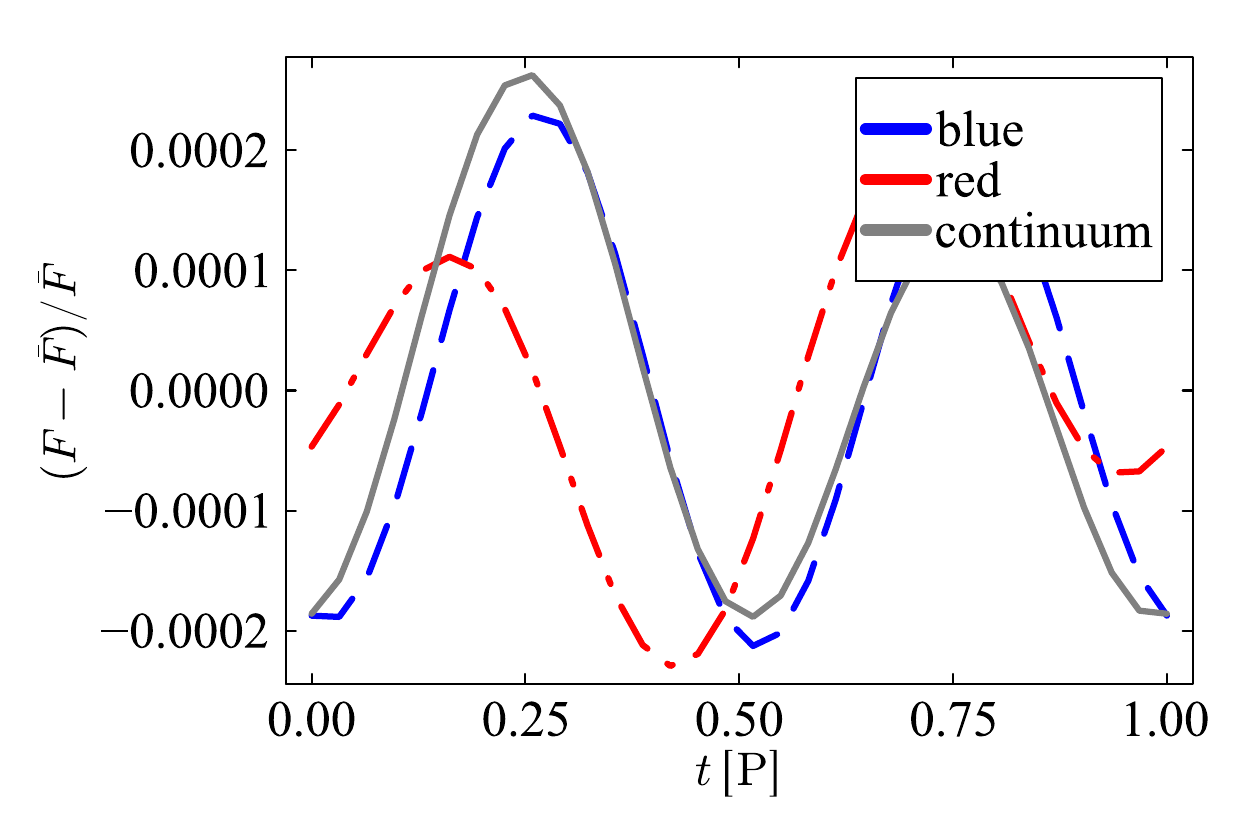}
    }
    \subfigure[Single MBH with the same continuum light curve as the binary case where $q=0.75$]{
    \includegraphics[width=7 cm]{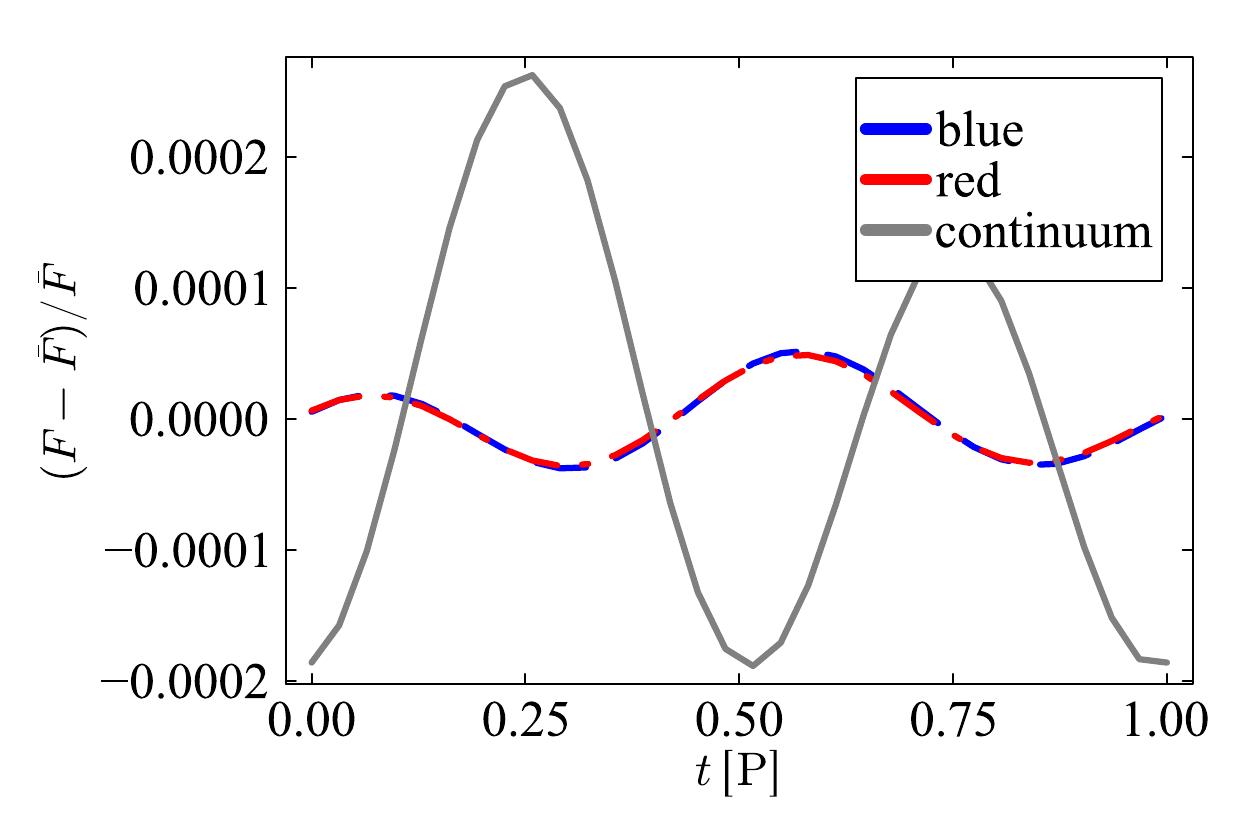}
    }
    \caption{Light curves from BELs for binaries with $a=10^{-4.2} \ \mathrm{pc}$, $M_1=10^6 \ \mathrm{M_{\odot}}$ and mass ratio $q=[0.25,0.55,0.75]$.}
    \label{fig:lightcurvesq}
\end{figure*}
Since the second-order effects are small, identifying such systems as binary candidates by the observation continuum light curves will be difficult in real scenarios, as the instrumental noise, as well as the intrinsic AGN variability, may hide the effect of the Doppler boost.  Figure \ref{fig:lightcurvesq} shows the light curves for binaries (left panels) with a separation $a=10^{-4.2} \ \rm{pc}$ with the primary MBH mass of $10^6 \mathrm{M_{\odot}}$ and mass ratios $q=[0.25,0.55,0.75]$ and compares them with the BLR response to a single MBH showing the same continuum light curve (right panels). The y-axis range changes to show smaller and smaller relative variations for increasing values of q.

When the light curves resemble the example shown in the middle and lower rows of Figure \ref{fig:lightcurvesq}, the presence of two very similar peaks in the light curves, introduced by second-order effects, could prevent an efficient time lag computation with PyCCF. In particular, in the presence of noise, as the mass ratio approaches unity, a higher S/N would be required to correctly compute the time delays, making the proposed test more observationally challenging. A critical scenario is found when an equal-mass binary is considered. In this case, to first order, the continuum light curve will not be variable, and the only variability that will be visible in the absence of noise is the one induced by the second-order effects. Considering this variability, the binary and single black hole scenarios are almost indistinguishable. 
 
More detailed analyses of the time evolution of the whole BEL (and of a varying continuum) might still identify the presence of a binary. In the following, we propose a specific test for the limiting $q=1$ case in the case of axisymmetric BLRs (i.e. in the absence of a significant spiral pattern in the BLR emissivity).

\begin{figure}[h!]
    \centering
    \subfigure{
    \includegraphics[width=8 cm]{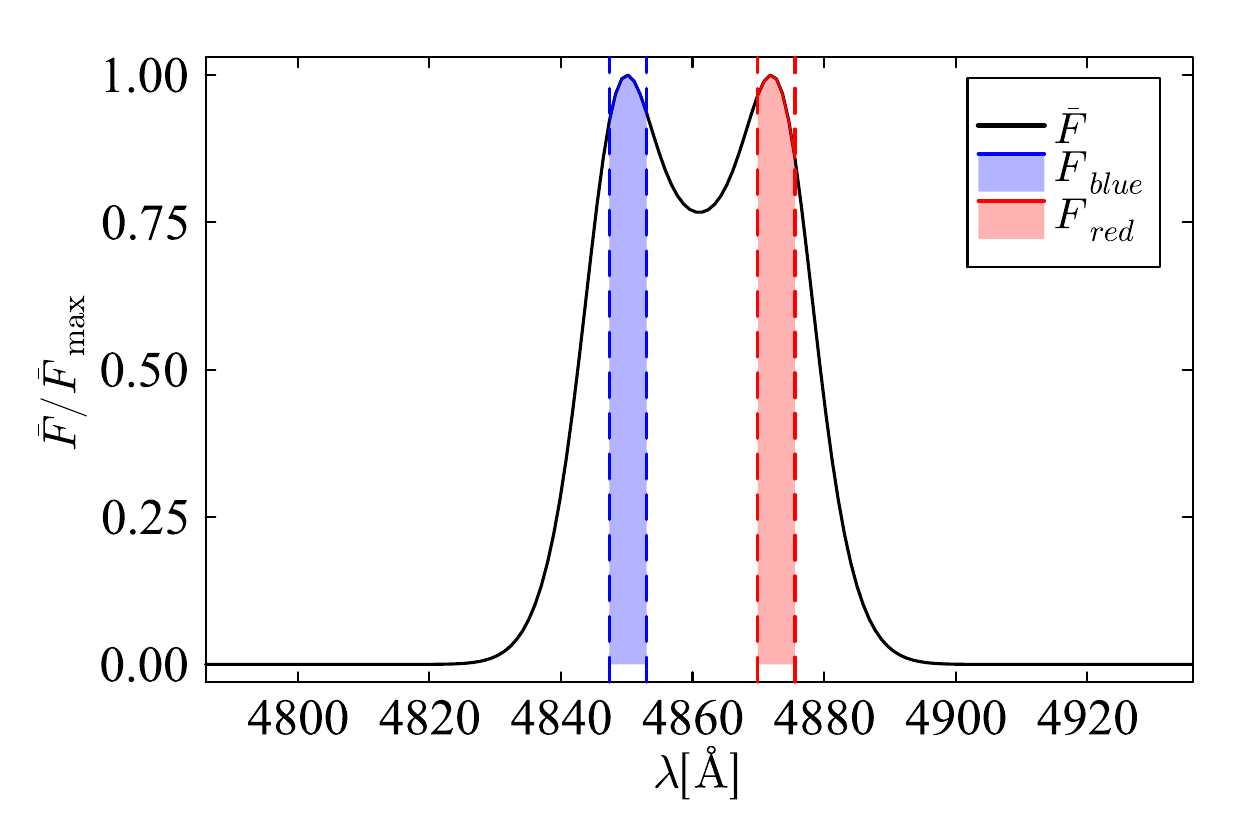}
    }
    \subfigure{
    \includegraphics[width=5.5 cm]{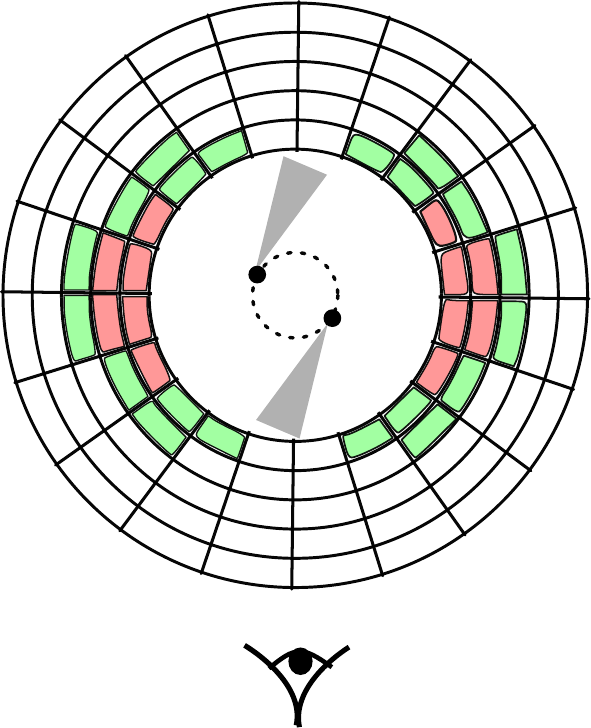}
    }
    \caption{Upper panel: Example of a cut in a different region of the BEL that is used to highlight the presence of a binary in the equal-mass case. Lower panel: qualitative example of the elements to be considered to highlight the presence of a binary in the case of an equal-mass binary with no spiral pattern. The red elements are the ones that are used in the method proposed for unequal-mass binaries, while the green ones are the ones that will encode the binary signature for the equal-mass binary.}
    \label{fig:qualitative_cut_equal_mass}
\end{figure}

\begin{figure}[h!]
    \centering
    
    \includegraphics[width= 8 cm]{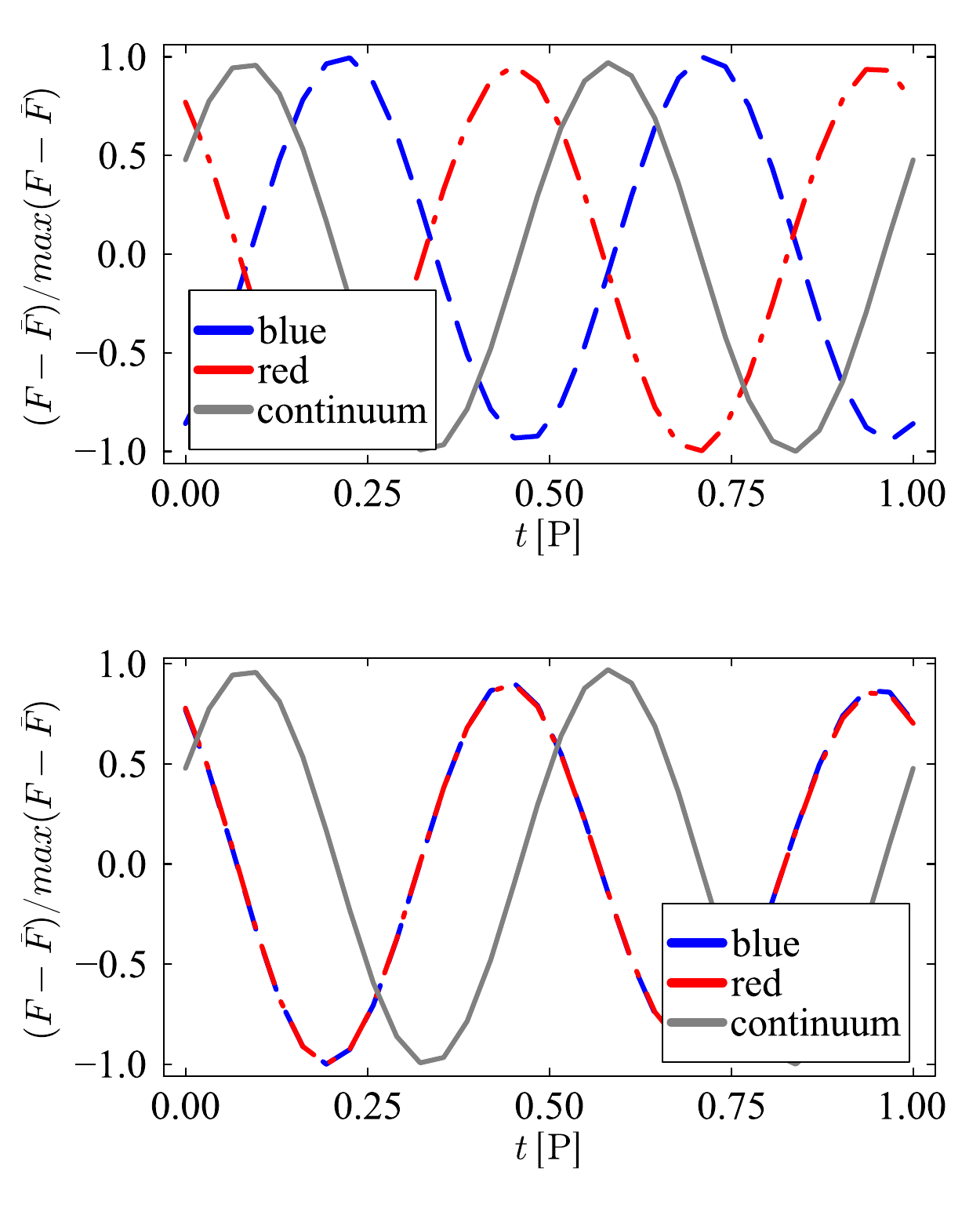}
    \caption{Example of how a cut in a different region of the BEL (those highlighted in figure~\ref{fig:qualitative_cut_equal_mass}) can highlight the presence of a binary with the red and blue light curves separated by a half-period lag.}
    \label{fig:continuuves_equal_right}
\end{figure}

The test can, however, be adapted for equal mass binaries in the specific case of symmetric BLRs. Consider, for example, two opposite elements of the BLR at the same distance $R_{\mathrm{el}}$ from the central source misaligned with respect to the line of nodes, as shown in the lower panel of figure~\ref{fig:qualitative_cut_equal_mass}. In the binary scenario, when one of the two MBHs is illuminating the farthest approaching element of the BLR, the other black hole will be illuminating the nearest receding element, inducing different delays between the red and blue light curves relative to the ionising continuum.
In the single MBH scenario, where the light is not beamed, no delay between the red and blue light curves is expected. An example of the resulting light curves obtained selecting the BEL wavelength range highlighted in the upper panel of  figure~\ref{fig:qualitative_cut_equal_mass} is shown in figure \ref{fig:continuuves_equal_right}. 

Considering the application of the test to real observations, we suggest that the test should be performed by cutting the BEL in different regions, as the binary mass ratio $q$ is not known a priori, and the choice of the region where the line should be cut is not straightforward. Considering different BEL regions, the light curves built considering the tails are expected to be more sensitive to unequal mass binaries, while the light curves built using a cut similar to the readapted test described above are expected to be more sensitive to equal mass binaries.

Another point that is worth stressing is that testing the presence of a binary by cutting the line as discussed in section \ref{subsec:light_curves} works particularly well for the BLR size we assumed (see section \ref{subsec:BLR}). When the inner and outer radii of the BLR are changed, the fraction of the maximum flux at which we find the limiting wavelength of the cut can be different from $\Bar{F}_{\rm{max \ (red/blue)}}/2$. For example, consider the case in which the outer radius of the BLR is bigger than $1.5 \ R_{\rm{BLR}}$. In this scenario, we expect more BLR elements to contribute to the BEL flux near the rest-frame wavelength. Cutting the line as in \ref{subsec:light_curves}, the resulting light curves are affected by elements with a lower line-of-sight velocity relative to the elements of interest. Considering unequal mass binaries, performing the test by cutting the BEL at different wavelengths can be beneficial in finding the regions that are mostly affected by elements with a high line-of-sight velocity. In particular, when the outer radius of the BLR is bigger than the one we assumed, the test can be performed multiple times cutting the line at wavelengths in which the BEL flux is a smaller and smaller fraction of $\Bar{F}_{\rm{max \ (red/blue)}}$ up until the considered region is not dominated by noise.

\end{appendix}
\end{document}